# Variable Planck's Constant: Treated As A Dynamical Field And Path Integral

Rand Dannenberg
Ventura College, Physics and Astronomy Department, Ventura CA
rdannenberg@vcccd.edu
rand@randtek.org

January 28, 2021

**Abstract.** The constant $\hbar$ is elevated to a dynamical field, coupling to other fields, and itself, through the Lagrangian density derivative terms. The spatial and temporal dependence of $\hbar$ falls directly out of the field equations themselves. Three solutions are found: a free field with a tadpole term; a standing-wave non-propagating mode; a non-oscillating non-propagating mode. The first two could be quantized. The third corresponds to a zero-momentum classical field that naturally decays spatially to a constant with no ad-hoc terms added to the Lagrangian. An attempt is made to calibrate the constants in the third solution based on experimental data. The three fields are referred to as actons. It is tentatively concluded that the acton origin coincides with a massive body, or point of infinite density, though is not mass dependent. An expression for the positional dependence of Planck's constant is derived from a field theory in this work that matches in functional form that of one derived from considerations of Local Position Invariance violation in GR in another paper by this author. Astrophysical and Cosmological interpretations are provided. A derivation is shown for how the integrand in the path integral exponent becomes $L_c/\hbar(r)$, where $L_c$ is the classical action. The path that makes stationary the integral in the exponent is termed the "dominant" path, and deviates from the classical path systematically due to the position dependence of $\hbar$. The meaning of variable $\hbar$ is seen to be related to the rate of time passage along each path increment. The changes resulting in the Euler-Lagrange equation, Newton's first and second laws, Newtonian gravity, Friedmann equation with a Cosmological Constant, and the impact on gravitational radiation for the dominant path are shown and discussed.

## 1.0 Introduction

There are three premier problems in physics, that Occam's Razor, as a guiding principle, suggests one seek a single explanation for: the nature of dark energy; the nature of dark matter; the nature of fundamental physical constants.

The idea studied is whether Planck's constant, a dimensioned quantity, may vary. The author uses the field $\psi$, where $\hbar=\beta\psi$. A comment on dimensioned versus dimensionless constants is given in Appendix 1.

What is the physical meaning of a variable Planck's constant? Could it ever be sensed by human beings? What conservation laws would remain in effect? What well-established theory would it conflict with? The goal here is to find the implications of this, examine how such an effect could be applied to the subject of the former paragraph, and find astrophysical and cosmological constraints on the variation that admit it, or rule it out. A logical course will be followed within the most basic formalisms, field theory, and the path integral. The authors hope is that the idea and results are memorialized, to be used in unforeseen ways, by others, either to support, or disprove, whether fundamental constants vary in any way. The author himself does not necessarily fully believe (or advocate) the main ideas presented in this paper, or even the information used as evidence for the variations of fundamental constants.

Fundamental constants are measured very carefully, seem to be only very weakly dependent on position, and if time varying, this time variation must be very slow. A solution is sought that is consistent with those observations that still might explain how a constant can come into being from some sort of field, that can describe energies associated with the constant, and variations with time or position, if any. One must also explain why the constants do not seem to be limited by the need to propagate at a speed $c$ or lower to have an effect – they are infinite in range, everywhere at all times, nearly equal in magnitude in all locations, persistent in duration, operate seemingly without any mechanism, and there are no easily observed particles associated with them. These observations are consistent with the non-propagating modes of fields, or static fields.

The following section is reproduced nearly *verbatim* from the authors own article: "Position Dependent Planck's Constant in a Frequency-Conserving Schrödinger Equation" *Symmetry* **2020**, *12*, 490; doi:10.3390/sym12040490. Some minor alterations in the text include a reordering of the same references, and a renumbering of the equations to be consistent with present article:

"The possibility of the variation of fundamental constants would impact all present physical theory, while all reported variations or interpretations of data concluding a constant has varied are extremely controversial. Examples of work in this area include Dirac's Large Number Hypotheses [1], the Oklo mine from which could be extracted a variation of the fine structure constant [2,3], and the observations of quasars bounding the variation of the latter per year to one part in $10^{17}$ [4-6]. Recent theoretical work includes the impact of time dependent stochastic fluctuations of Planck's constant [7], and the effect of a varying Planck's constant on mixed quantum states [8]. An authoritative review of the status of the variations of fundamental constants is given in [9].

Publicly available Global Positioning System (GPS) data was used to attempt to confirm the Local Position Invariance (LPI) of Planck's constant under General Relativity [10-11]. LPI is a concept from General Relativity, where all local non-gravitational experimental results in freely falling reference frames should be independent of the location that the experiment is performed in. That foundational rule should hold when the fundamental physical constants are not dependent on the location. If the fundamental constants vary universally, but their changes are only small locally, then it is the form of the physical laws that should be the same in all locations.

The LPI violation parameter due to variations in Planck's constant is called $\beta_h$. The fractional variation of Planck's constant is proportional to the gravitational potential difference and $\beta_h$. The value found in [10] for variations in Planck's constant was $|\beta_h|<0.007$.

and,

The study did not report on the altitude dependence of Planck's constant above the earth. A very recent study involving the Galileo satellites found that GR could explain the frequency shift of the onboard hydrogen maser clocks to within a factor of $(4.5\pm3.1)\times10^{-5}$ [12], improved over Gravity Probe A in 1976 of $\sim 1.4\times10^{-4}$, these are the $\alpha_{rs}$ redshift violation values that may be compared to $\beta_h$.

Consistent sinusoidal oscillations in the decay rate of a number of radioactive elements with periods of one year taken over a 20 year span has been reported [13-18]. These measurements were taken by six organizations on three continents. As both the strong and weak forces were involved in the decay processes, and might be explainable by oscillations of $\hbar$ influencing the probability of tunneling, an all electromagnetic experiment was conducted, designed specifically to be sensitive to Planck's constant variations [19]. Consistent systematic sinusoidal oscillations of the tunneling voltage of Esaki diodes with periods of one year were monitored for 941 days. The charge carrier tunneling of an Esaki diode can be modeled using the Wentzel–Kramers–Brillouin (WKB) approximation. The result is a tunneling probability that is an exponential function of the physical width of the barrier, and the applied voltage. The tunnel diode oscillations were attributed to the combined effect of changes in the WKB tunneling exponent going as $\hbar^{-1}$, and changes in the width of the barrier going as $\hbar^{2}$. The electromagnetic experiment voltage oscillations were correctly predicted to be 180 degrees out of phase with the radioactive decay oscillations. This data can be made available for independent analysis by requesting it from the author of [19].

It is both controversial, and easily criticized, to suspect that the oscillations of decay rates and tunnel diode voltage are related to the relative position of the sun to the orbiting earth, and that there are resulting oscillations in Planck's constant due to position dependent gravitational effects, or effects with proximity to the sun. It should be mentioned that there have been studies in which it was concluded there was no gravitational dependence to the decay rate oscillations [20-21]. There is also dispute in the literature concerning the reality of the decay rate oscillations [22-24]."

As further clarifications, the LPI violation parameter result for Planck's constant discussed above was taken by its investigator to be a null result, concluding it to be zero, because its variation was less than the measurement uncertainty with a bound about zero, [10-11], while [25] discusses a mechanism for why such violations may be undetectable in experiments with clocks and light. In the second paragraph of the above excerpt from [25], the term "universally" is used as an abbreviation for "temporal and spatial variations over cosmological scales", and "locally" was used to mean "temporal and spatial variations over scales of the everyday human experience of life on Earth", or "benchtop experiments". The intentional search for variations in Planck's constant seems to be limited to only to the references [10-11], and the diode experiment of [19]. In reviews of this paper, it is often pointed out that the decay rate oscillations cannot be used as direct evidence of variations in Planck's constant, and the author admits this in the article. However, reviewers, thus far, never comment on, or even acknowledge, the diode test [19], yet then readily posit that there is no evidence for Planck's constant variations.

A new paper on positional (gravitational) variations in the fine structure constant from spectral analysis of Fe V absorption of the white dwarf G191-B2B reports that the fine structure constant increases in high gravitational fields [26], building on the earlier work of Berengut [27]. These papers are part of a very limited body of literature experimentally examining the positional variation of a fundamental constant, another is [28] on the proton-electron mass ratio variation about white dwarfs GD133, and G29-38. An essentially null result for variations of the fine structure constant on the Earth in its orbit around the sun is found in [29], in disagreement with [19] on Planck's constant on the Earth in orbit about the sun. None of these results approach a significance level of $6\sigma$, and in all, unaccounted for systematic effects may be influencing the results. However, this is an active and ongoing area of research. Workers using the HST STIS, the ESPRESSO instrument at the VLT, the Keck telescope, etc. (this author is one of them, on the experimental-spectroscopic side) are continuing to look for variations in the fine structure constant, refining analysis methods, eliminating systematics, and comparing experimental results to those predicted by classical field theory using various coupling setups to gravity. Since there is presently no undisputed (no accepted) data concerning such variations, it is presently not known whether constants are dynamical fields, or whether they are coupled to gravity. This paper is mostly concerned with spacetimes that are "sufficiently flat" in approximation, where gravity does not directly enter, explained in Section 2.9.

## 2.0 Dynamical Fields

At this time, it is not known conclusively whether the variation of a constant signifies that it is assuredly a dynamical field, or not, or is something else entirely, not yet understood. In a separate paper by this author, issues specific to the Schrödinger equation in a single-particle, non-relativistic, non-field theoretic framework for a position-dependent $\hbar$, that is not treated as a dynamical field, were examined [25], the reference of the excerpted section. That work is relevant to the present paper for two reasons. The first reason is that the positional variation of $\hbar$ is derived in [25] from a completely different starting point, that results in the same in functional form as one derived in this paper.

In this paper's Section 2, variations in $\hbar$ will be treated as a scalar dynamical field, coupling fields through the derivative terms in the Lagrangian density, but not coupled to gravity. The second reason is that one of the results to follow in this paper suggests that frequency may be a more fundamental parameter than energy, and the Schrödinger equation of [25] is frequency-conserving and Hermitian. Continuing the modified excerpt from [25]:

"The scope of much work with variable constants as dynamical fields has been to address unsolved problems in cosmology. Consider the Cosmon of Wetterich [30-31], using a field dependent prefactor to the derivative term in the scalar Lagrangian that decays to a constant value for large field values, acting like variable Planck mass [31]. Existing theories with varying constants as fields are the Jordan-Brans-Dicke scalar-tensor theory developed in the late 1950's with variable $G$, and Bekenstein variable fine structure constant theory developed in 1982 [32-33]. The latter did not contain gravity, and was concerned with the electromagnetic sector. Albrecht, Magueijo and Moffat, examined a variable $c^4$ to attempt to explain the flatness and horizon problems, the cosmological constant problem, and homogeneity problem [34]. Cosmologies of varying $c$ were examined by Barrow [35] and Moffat [36-37].

$$S_{GR}=\int\left\{\begin{array}{l}\dfrac{(c_o\mathbb{C})^4}{16G_o\pi}\xi R+\\ \dfrac{(\hbar_o\psi)^2}{2}g^{\mu\nu}\nabla_\mu\psi\nabla_\nu\psi+\dfrac{(\hbar_o\psi)^2}{2}\dfrac{w}{\xi}g^{uv}\nabla_\mu\xi\nabla_\nu\xi+\dfrac{(\hbar_o\psi)^2}{2}g^{\mu\nu}\nabla_\mu\mathbb{C}\nabla_\nu\mathbb{C}\\ +\lambda\xi\psi\mathbb{C}R+L_m\{\hbar,c,G\}\end{array}\right\}\sqrt{-g}\,d^4x \tag{1}$$

$$\frac{\partial}{\partial x_o}=\frac{1}{c_o\mathbb{C}}\frac{\partial}{\partial t} \tag{2}$$

$$G=G_o/\xi \tag{3}$$

$$\hbar=\hbar_o\psi \tag{4}$$

$$c=c_o\mathbb{C} \tag{5}$$

Equations (1) to (5) show in a single form an amalgam of possible couplings including a Jordan-Brans-Dicke-like scalar-tensor theory of alternative General Relativity with variable $G$, an Albrecht-Magueijo-Barrow-Moffat-like field for $c$, a field for $\hbar$ like that of [32-33], which is different than the form of Bekenstein's for variable $e^2$ whose representative field squared divided the derivative terms. There is also the field theory of Modified Gravity (MOG) of Moffat, and the Tensor-Vector-Scalar (TeVeS) gravity of Bekenstein. There are many ways all the constants might be represented as fields, and many ways they might be coupled. Coupling fields together in this way is the accepted approach for the treatment of a constant."

The scalar field setup used for $\hbar$ will couple through the derivative terms, and to itself. Cosmological and astrophysical interpretations of this setup will be developed. The fields in the prefactors to follow are literally assigned to represent Planck's constant, distinct from $c$, or the fine structure constant, and $e$ is not present.

Consider these six specifics concerning the mathematics of scalar fields:

i. *k*-Essence Theory

These models where introduced as an alternative to constant dark energy, to explain the accelerated expansion of the universe, and the time evolution of the cosmological constant [38-42]. The actions involve $\mathcal{L}_k = F$ for any function the kinetic terms $X$ and scalar field $\psi$, usually minimally coupled to gravity,

$$\mathcal{L}_k = F(\psi, X) \tag{6}$$

$$X = \frac{1}{2} g^{\mu\nu} \nabla_\mu \psi \nabla_\nu \psi \tag{7}$$

Thus, the *k*-Essence models may encompass the field to be studied here that is linear in $X$, and there may already be a well-studied mathematical solution that has not been directly called out as $\hbar$, or any other familiar constant, other than the cosmological constant.

ii. Field Redefinition

It is possible to redefine a coupled scalar field so that the kinetic terms look like that of a normal free-field theory. This is most easily accomplished when the coupling of $\psi$ is not to the kinetic terms of another field (and in some cases when it is), and when $\mathcal{L}_k$ is linear in $X$. When $\psi$ is redefined, new interaction terms then appear in the Lagrangian, and a Hamiltonian may be derived from the Euler-Lagrange equations with respect to the new field. The perturbative expansion of QFT may then be used for the theory, and the *S*-matrix is not altered by the redefinition. Thus, amplitudes for various processes may be computed, despite the complexity of the original Lagrangian, with alterations in the interpretations of interaction terms and pre-factors.

iii. Canonical Quantization

The standard prescription will be applied to the most basic Klein-Gordon free-field. A variation is offered, pertaining to an infinite number of field redefinitions that the particular setup under study is amenable to.

iv. Non-Propagating Modes and Static Modes

Non-propagating modes, and also static fields, are a reasonable candidate to represent a fundamental constant, as they do not require the propagation of particles to have a remote effect. The non-propagating modes involve particles only when the their values change, and readjust to give the new, static field.

v. Vacuum Solutions

The classical vacuum solutions are well known, and here are linked to non-propagating modes. Vacuum solutions were examined by the author, in which Planck's constant appears as an unnamed field in the denominator of a Lagrangian density, the field called $\varphi$ in Equations (44) and (45) of [43], sketched initially in the unpublished paper [44]. The latter paper is not yet complete.

vi. Tadpoles and Lagrangians with Linear Field Dependences

The coupling setup will produce a linear field interaction term (tadpole) from the mass term, inducing a non-zero vacuum expectation value (VEV). The standard treatment is to eliminate the tadpole with a field redefinition, so the that *S*-matrix elements can be more easily computed by normal means. Retaining the linear term, however, spoils the interpretations of mass and propagation. Here, a divergent, static classical solution will prompt the mass to be set to zero, for four reasons: the boundary condition assumed is that the field should not diverge far from the origin; the consensus is that any varying constant would be described by a nearly massless scalar field; the mathematics is easier; terms and coefficients retain their interpretations. The tadpole may have other consequences for Planck's constant and coupling, but that will be relegated to further work.

There are a great number of rather exotic fields under study. Some have been intentionally linked to varying constants at the outset, per those in the introduction. Of those fields that have *not* already been linked to the variation of a familiar constant from the start, one may wonder if some theories might be, after the fact. Identifying them would be no small undertaking.

One objective of this effort is to publicize that mathematical solutions may already exist for fields that could represent fundamental constants, but have not been brought to light, either because of the reticence to talk about the variation of a constant, or the field was not recognized as possibly representing a constant. There is also a tendency in field theory to work in natural units, $c = G = \hbar = 1$, and this would tend to suppress thinking about fields as representing, and evolving into, the fundamental constants with which we are most familiar. In cosmological theories, a field capable of explaining many observations, left unrecognized as a constant, inadvertently or intentionally, reduces the possibility of experimentally substantiating the theory, as it reduces the scope of experiments that might be capable of detecting its implications.

Per Occam's Razor, it is reasonable to start with theories that involve varying fields that are intentionally representative of the fundamental constants most familiar to us, reduced to the utmost of simplicity. The one that has been given minimal treatment is $\hbar$, now treated below, for the least complicated case.

The paper outline is as follows: 1) The Lagrangians of importance will be developed; 2) Propagating and non-propagating mode solutions will be examined, and where possible, quantized; 3) Cosmological and astrophysical interpretations will be offered; 4) The path integral will be developed for a position dependent Planck's constant; 5) Relationships of the latter to dark matter, the cosmological constant, and gravitational waves will be presented.

**2.1 Classical Field Lagrangian Densities**

Consider two symmetrically coupled scalar Klein-Gordon fields of mass $m$ and $M$ with coupling constant $g$, showing the unburied location of the constants $\hbar$ and $c$ in the Lagrangian density $\mathcal{L}$,

$$\mathcal{L} = \frac{1}{2}\hbar^2 \partial_u \varphi \partial^u \varphi - \frac{1}{2} m^2 c^2 \varphi^2 + \frac{1}{2}\hbar^2 \partial_u \psi \partial^u \psi - \frac{1}{2} M^2 c^2 \psi^2 - g\psi\varphi \tag{8}$$

The Lagrangian density of Equation (8) produces the normal coupled Klein-Gordon equations for the fields (or particles) $\psi$ and $\varphi$. The constant $\hbar$ might be associated with the fields directly (though this does not matter if $\hbar$ is just a normal constant). If so, as a prelude to making $\hbar$ a field, it is shown as operated on by the derivatives,

$$\mathcal{L} = \frac{1}{2} \partial_u \hbar \varphi \partial^u \hbar \varphi - \frac{1}{2} m^2 c^2 \varphi^2 + \frac{1}{2} \partial_u \hbar \psi \partial^u \hbar \psi - \frac{1}{2} M^2 c^2 \psi^2 - g\psi\varphi \tag{9}$$

Whether the latter is necessary or not is an unknown, but is an option to investigate, and there are many possible coupling schemes, other than the one to be discussed.

Let it now be supposed that $\hbar$ is itself a dynamical field, which would be a natural way to introduce it as a non-constant. If $\hbar$ is to have the usual units of J·s, the unit keeping then necessitates the introduction of yet another constant $\beta$, where $\hbar=\beta\psi$.

Selection of a Lagrangian is per Occam's Razor, and is the simplest form that combines $\hbar$ as its own pre-factor, and as the pre-factor of all other to-be-quantized fields.

From (8),

$$\mathcal{L} = \frac{1}{2}(\beta\psi)^2 \partial_u \varphi \partial^u \varphi - \frac{1}{2} m^2 c^2 \varphi^2 + \frac{1}{2}(\beta\psi)^2 \partial_u \psi \partial^u \psi - \frac{1}{2} M^2 c^2 \psi^2 - g\psi\varphi \tag{10}$$

The key idea here is that the quantum mechanical field for $\hbar$ also functions as its own $\hbar$. To coin a term, the field for $\hbar$ is its own "actionizer", and acts on all other to-be-quantized fields with the same proportionality constant $\beta$. The fields for $\hbar$ are dubbed the "acton".

### 2.2 Planck's Constant as a Self-Actionizing Field in Isolation

Consider first a much more standard form of coupling. Take Equation (8) with $M$=$m$=0, describing two massless fields interacting through a coupling term that is separate from the derivative term. Then let the two fields be the same field, that is, let $\varphi=\psi$. Equation (8) then becomes, with $\hbar$ a bona fide constant,

$$\mathcal{L}=\mathcal{L}_k=F\left(\psi,X\right)=\hbar^2\partial_u\psi\partial^u\psi-g\psi^2 \tag{11}$$

The resulting equation of motion is,

$$\partial_t^2\psi-c^2\nabla^2\psi+\frac{gc^4}{\hbar^2}\psi=0 \tag{12}$$

From Equation (11) and (12) it is seen that a fields ability to couple to itself separately from the derivative term, that is it self-interacts, produces the equivalent of a mass equal to $m^2 = g$. Such setups are normally avoided, so as to not cause confusion with actual particle masses, who's values and mass ratios are fundamental constants.

Quadratic self-interaction terms are interpreted as mass in a free field. In Lagrangians with additional interaction terms, there must be no linear terms if the quadratic terms are to be interpreted as a physical mass, even though the theory can still be solved. Such terms are important in symmetry breaking, but here they will not be treated.

### 2.3 Infinitely Repeated Field Redefinition $\chi=\psi^2$ and Relation to Vacuum Energy

Now promote $\hbar$ to a field $\beta\psi$, then, redefine the field. So that it comes as no surprise to the reader, an interpretation of these results will be given Section 2.8, relating them to the cosmological constant/dark energy, and vacuum energy. A variant of this idea will also be developed in section 3.5.2 in the path integral treatment. The central idea is that the energy that sustains (or is associated with) a fundamental constant, in the absence of particles, is the vacuum.

The Lagrangian densities, equations of motion, and Hamiltonian for $\psi$ and its redefinition $\chi$=$\psi^2$ are,

$$\mathcal{L}_\psi=\frac{1}{2}\left(\beta\psi\right)^2\partial_u\psi\partial^u\psi-\frac{1}{2}M^2c^2\psi^2 \tag{13}$$

$$\ddot{\psi}-\nabla^2\psi+\frac{M^2c^2}{\beta\psi}+\frac{\dot{\psi}^2}{\psi}=0 \tag{14}$$

$$\mathcal{H}_\psi=\frac{1}{2}\left(\beta\psi\right)^2\dot{\psi}^2+\frac{1}{2}\left(\beta\psi\right)^2\left(\nabla\psi\right)^2+\frac{1}{2}M^2c^2\psi^2 \tag{15}$$

$$\mathcal{L}_\chi = \frac{1}{8}\beta^2\partial_u\chi\partial^u\chi - \frac{1}{2}M^2c^2\chi \tag{16}$$

$$\ddot{\chi} - \nabla^2\chi + 4\frac{M^2c^2}{\beta^2} = 0 \tag{17}$$

$$\mathcal{H}_\chi = \underbrace{\frac{1}{8}\beta^2\dot{\chi}^2 + \frac{1}{8}\beta^2(\nabla\chi)^2}_{\mathcal{H}_\chi^f} + \underbrace{\frac{1}{2}M^2c^2\chi}_{\mathcal{H}_\chi^i} \tag{18}$$

Note that the mass terms of $\mathcal{L}_\chi$ and $\mathcal{H}_\chi$ , (16) and (18) linear in $\chi$, the tadpole. The free-field term and tadpole term have been named in the underbrackets of (18). The tadpole term no longer physically represents a mass, rather an induced non-zero vacuum expectation value (VEV). If one wishes to use $\chi$ to compute $S$-matrix elements, the vacuum expectation value must vanish, either by setting $M$=0, or by redefining the field again, per $\chi \rightarrow \chi\text{-}\chi^{(\mathrm{VEV})}$. If the latter redefinition were utilized in a coupling involving more fields than $\chi$, then all the coefficients and interaction terms in the Lagrangian would also be changed from their original interpretations, so that will be avoided.

Even more concretely, $\chi$ is not in the mass term of the equation of motion (17). As a result, the static *classical* solution of the equation of motion for $\chi$ diverges as $r^2$. The non-propagating modes are, here, more heavily weighted candidates to represent the familiar constants. Concerning the boundary conditions for fields that may represent physical constants, it is sensible that they should not diverge with increasing range.

For all of the reasons mentioned here and in the prior sections, Occam's Razor leads to the additional simplification that $M$=0, and the massless Klein-Gordon equation if $\chi \rightarrow 2\chi$ (although the latter will not be employed). $\hbar^2 = \beta^2\chi$ is a viable solution, provided the mass is zero, to avoid a classical solution that diverges with increasing range. This finding is at least consistent with the consensus that the masses of "fundamental constants fields" be near zero [9].

Note that the massless Klein-Gordon equation (17) with $M$=0 for the squared field $\hbar^2 = \beta^2\chi$ is equivalent to (14) with $M = 0$ for the unsquared field $\hbar = \beta\psi$. That is, the equation of motion for Planck's constant $\hbar = \beta\psi$ is not just a massless Klein-Gordon equation for $\psi$, because the field must support itself. These are important results, arrived at because the field was specifically called out as representing $\hbar$ at the beginning. Were $M$ not zero, one of the cosmological interpretations of this field, discussed in Section 2.7.1, would not be possible, and the averaging scheme in Appendix 3, pertaining to the non-propagating mode solution of Section 2.7, would not produce a sensible result.

Quantization of $\chi$ follows by the standard procedure, with Fourier expansion of the fields, making Fourier coefficients operators,

$$E_{\chi} = \frac{1}{4}\beta^2 \int \frac{d^3 p}{(2\pi)^3} \frac{\omega_{\overline{p}}}{c} \left( a_{\overline{p}}^{\dagger} a_{\overline{p}} + \frac{(2\pi)^3}{2} \delta^{(3)}(0) \right) \tag{19}$$

The vacuum term has been left intentionally in (19), and the divergent term is usually explained by two singularities that arise in the treatment, the infrared and ultraviolet divergences.

It is now natural to ask whether $\beta$, the new constant, should also be represented by a field. What is interesting about this setup, is that it is amenable to repeated application of the same redefinition procedure. One finds for the $n^{\text{th}}$ repeated application,

$$\mathrm{L}_{\psi}^{(n)} = \frac{1}{2}\frac{1}{4^n}\left(\beta^{(n)}\right)^2 \partial_u \psi^{(n+1)} \partial^u \psi^{(n+1)} \tag{20}$$

$$\beta^{(n-1)} = \beta^{(n)} \psi^{(n)} \tag{21}$$

$$\psi^{(n)} = \left(\psi^{(n-1)}\right)^2 = \left(\psi^{(1)}\right)^{2n} \tag{22}$$

$$\hbar = \beta^{(n)} \prod_{i=1}^{n} \psi^{(i)} \tag{23}$$

$$E_{\chi}^{(n+1)} = \frac{1}{4^n}\left(\beta^{(n)}\right)^2 \int \frac{d^3 p}{(2\pi)^3} \frac{\omega_{\overline{p}}^{(n+1)}}{c} \left( a_{\overline{p}}^{(n+1)\dagger} a_{\overline{p}}^{(n+1)} + \left( \frac{(2\pi)^3}{2} \delta^{(3)}(0) \right)^{(n+1)} \right) \tag{24}$$

Normal ordering would remove the vacuum term, and it would be given no further thought, though normal ordering is the equivalent of a subtraction of it - or just throwing it out. Intentionally leaving it, for $n=\infty$, for a finite number of particles, the term of (24) that has a clearer prospect of remaining non-zero is the vacuum term, written in (25). Without a formal investigation of the infinities, one may tentatively conclude that the energy that sustains Planck's constant is that of the vacuum,

$$E^{(\infty)} = \frac{1}{4^{\infty}}\left(\beta^{(\infty)}\right)^2 \int \frac{d^3 p}{(2\pi)^3} \frac{\omega_{\overline{p}}^{(\infty)}}{c} \left( \frac{(2\pi)^3}{2} \delta^{(3)}(0) \right)^{(\infty)} = E^{(1)} \tag{25}$$

Further, the redefined field should have the same energy as before redefinition. Per (25), this would be possible if there are no permitted excitations of the field, that is, no particles at all, and all of the operators are zero, namely, static, and non-propagating fields.

While the argument is not rigorous, it does hint at the importance of the non-propagating modes, vacuum solutions, solutions with zero momentum, static solutions, and why such excitations are not in the common realm of experience. Despite that this manipulation may seem to be an overcomplication of the simplest solution available, it is shown for a reason: what constants actually are is not understood, and results (24) and (25) involving the infinity plus infinite redefinition is an unturned stone, mathematically.

In (25), the only variable associated directly with the field/particle in the integrand is the frequency. In this light, one may speculate whether the most fundamental physical parameter describing the vacuum is the frequency. This was the result mentioned in the introduction that prompted the investigations made of a frequency-conserving form of the Schrödinger equation in [25].

The discussion will continue with only the first field redefinition. The relationship between Planck's constant and the energy of the vacuum will be taken up again in Sections 2.8 and 3.5.2.

### 2.4 Propagating and Non-Propagating Classical and Quantized Solutions for $\chi$

Perhaps the reason there is no propagation required for a constant to have an effect is because the field associated with it has zero momentum, that is, the occupation number of the particles is zero, and the modes are non-propagating, leaving only the vacuum. From the equation of motion for $\chi$ with $M$=0, one finds by separation of variables, and completely standard canonical quantization,

$$\chi = S_{\chi}(\overline{r})\varphi_{\chi}(x_o) \tag{26}$$

$$\frac{\ddot{\varphi}_{\chi}}{\varphi_{\chi}} = \frac{\nabla^2 S_{\chi}}{S_{\chi}} = -\frac{\omega_{\overline{p}}^2}{c^2} = -\overline{p}\cdot\overline{p} = -p^2 \tag{27}$$

$$\varphi_{\overline{p}\chi} = A_{\overline{p}} e^{i\frac{\omega_{\overline{p}}}{c}x_o} + B_{\overline{p}} e^{-i\frac{\omega_{\overline{p}}}{c}x_o} \tag{28}$$

$$S_{\overline{p}\chi} = C_{\overline{p}} e^{i\overline{p}\cdot\overline{r}} + D_{\overline{p}} e^{-i\overline{p}\cdot\overline{r}} \tag{29}$$

$$D_{\overline{p}} = A_{\overline{p}} = 0 \tag{30}$$

$$\chi_{\overline{p}} = C_{\overline{p}} B_{\overline{p}} e^{i\overline{p}\cdot\overline{r} - i\frac{\omega_{\overline{p}}}{c}x_o} + c.c. = \tilde{a}_{\overline{p}} e^{i\overline{p}\cdot\overline{r} - i\frac{\omega_{\overline{p}}}{c}x_o} + \tilde{a}_{\overline{p}}^{*} e^{-i\overline{p}\cdot\overline{r} + i\frac{\omega_{\overline{p}}}{c}x_o} \tag{31}$$

$$\chi = \int \frac{d^3 p}{(2\pi)^3} \sqrt{\frac{c}{8\omega_{\overline{p}}}} \left( a_{\overline{p}} e^{i\overline{p}\cdot\overline{r}} + a_{\overline{p}}^{\dagger} e^{-i\overline{p}\cdot\overline{r}} \right) \tag{32}$$

$$\lim_{\overline{p}\to 0} \chi_{\overline{p}} = \chi_o = (\tilde{a}_o + \tilde{a}_o^{*}) \tag{33}$$

The equation of motion for $\chi$ is linear and homogeneous, allowing the sum over the solutions $\chi_p$. In taking the limit $p \to 0$ *after* finding the solutions, it is seen that the zeroth component $\chi_o$ is constant, so $\hbar^2 = \beta^2\chi_o$ is constant and real. The equations (27) to (33) are actually a special case of a more general solution, and equations (26) and (27) are compared to it Appendix 2, and is pertinent to quantization of a standing wave field.

### 2.5 Non-Propagating Classical Solutions for $\chi$ and $\psi$

Consider taking the limit $p \to 0$ *before* solving the differential equation as a vacuum (non-propagating) solution. One finds a different solution. Mathematically, it is known the limit of the solution will not equal the solution of the limit. Yet, there is no reason physically to exclude either as a solution in the interest of finding new explanations for variations of constants, which are not well-understood. Such spatial solutions are those of Laplace's equation, the same as the equation of the Newtonian gravitational potential in vacuum, or the electric potential in vacuum. The solutions for the zero-momentum field representing $\hbar^2$ are therefore the allowed classical vacuum solutions. From (27) with first taking $p \to 0$,

$$\underline{S}_{o\chi}(\bar{r}) = \sum_{l=0}^{\infty}\sum_{m=0}^{\infty}\underline{S}_{o\chi}(\bar{r})\Big|_{l}^{m} = \sum_{l=0}^{\infty}\sum_{m=0}^{\infty}\left(A_l r^{l} + B_l r^{-l-1}\right)P_l^m(\cos\phi)\left(S_m\sin(m\theta) + C_m\cos(m\theta)\right) \tag{34}$$

$$\underline{\varphi}_{o\chi}(x_o) = b_1 + b_2 x_o \tag{35}$$

where the underline means the limit $p \to 0$ is taken before solving the differential equation, and with the spatial form selected so as to be real. The sub- and superscripts are the azimuthal and polar indices of the solution. Of particular interest is the solution for $l=m=0$ with all other coefficients zero,

$$\underline{S}_{o\chi}(\bar{r})\Big|_{l}^{m} = b_3 + \frac{b_4}{r} \tag{36}$$

$$\underline{S}_{o\chi}(\bar{r})\Big|_{0}^{0} = b_3 + \frac{b_4}{r} \tag{36}$$

Note that in (35) and (36) there are constants $b_{1,2,3,4}$. The first and second pertain to the time dependence (35), and the third and fourth pertain to the spatial dependence of the $l=m=0$ solution of (34), renamed for neatness.

The reason for the interest is that (36) may be always positive, real, and decays to a constant at $r=\infty$. This way, very far from the origin of the acton, which is to represent a physical constant, there is no asymmetry of behavior in space. The solutions (34) to (35) could not be readily quantized by this author. There is, however, now a classical energy associated with the solution (35) and (36), despite that momentum is zero. The Hamiltonian density is from (18), (35) and (36),

$$\underline{H}_{o\chi}\Big|_{0}^{0} = \frac{\beta^2}{8}\left(\left(b_2 b_3 + \frac{b_2 b_4}{r}\right)^2 + \frac{\left(b_1 b_4 + b_2 b_4 x_o\right)^2}{r^4}\right) \tag{37}$$

and Planck's constant becomes,

$$\underline{\hbar}_{o\chi}\Big|_{0}^{0} = \beta\sqrt{\left(b_1 + b_2 x_o\right)\left(b_3 + \frac{b_4}{r}\right)} \tag{38}$$

It can be seen from (37) and (38) that Planck's constant, per this model, can change as a function of time and position, and, that there is an energy density associated with its existence that also changes as a function of time and position. From (34) there are many other profiles that might be explored.

Consider now another related solution path, where the field is not representative of $\hbar^2$, but of $\hbar$. The equation of motion for $\psi$ is now derived from Equation (13), and not its square $\chi$. The non-linear homogeneous equation of motion is (14).

Equation (14) is separable ($x_o$ dependent) if and only if $M$=0, another cause for this choice. Writing $\psi=S_\psi(r)\varphi_\psi(x_o)$, the equation of motion becomes, for a specific wavenumber $p$ or frequency $\omega_p$,

$$\frac{\ddot{\varphi}_\psi}{\varphi_\psi}+\frac{\dot{\varphi}_\psi^2}{\varphi_\psi^2}=\frac{\nabla^2 S_\psi}{S_\psi}+\frac{\left(\nabla S_\psi\right)^2}{S_\psi^{\ 2}}=-\frac{\omega_{\bar{p}}^{\ 2}}{c^2}=-p^2 \tag{39}$$

The limit $p\rightarrow 0$ is again taken first, now stipulating a radial dependence only, so using the radial term of the spherical Laplacian,

$$\underline{S}_{o\psi}(r)=\left(b_3+\frac{b_4}{r}\right)^{1/2} \tag{40}$$

$$\underline{\varphi}_{o\psi}(x_o)=\left(b_1+b_2x_o\right)^{1/2} \tag{41}$$

Squaring (40) and (41), then comparing to (34) and (35), one sees they are identical. Enforcing a radial dependence in $\underline{\psi}$ produces the same solution as $l$=$m$=0 for $\chi$=$\psi^2$. One finds,

$$\underline{\psi}_o=\sqrt{\underline{\chi}_o}\,\big|_0^0 \tag{42}$$

$$\underline{\hbar}_{o\psi}=\underline{\hbar}_{o\chi}\,\big|_0^0 \tag{43}$$

$$\underline{H}_{o\psi}=\underline{H}_{o\chi}\,\big|_0^0 \tag{44}$$

which is a reassuring check, showing that both representative fields are equivalent in this case. The Hamiltonian density used for (44) was (18) with $M$=0.

### 2.6 Another Non-Propagating Solution for $\chi$ and $\psi$ – Standing Waves

The following is in keeping with finding solutions that far from the acton origin do not show a spatial asymmetry. This is accomplished from (39) by stipulating that the solutions $\psi$ have spherical symmetry with only a radial momentum $p_r$, so $\psi$ has no angular dependence. The step produces localized standing wave solutions that do not decay in time. There are no resulting dot products of $p$ with position $r$ in the solutions, only the scalar product $p_r r$.

Continuing, the solutions to the non-linear equation of motion $\psi_p$ will then be individually squared to produce $\psi_p{}^2$, analogous to $\chi_p$ which one recalls is derived from a linear and homogeneous equation of motion, allowing solution summation. The $\psi_p{}^2$ will then be summed to form $\psi^2$, followed by quantization of $\psi^2$, not $\psi$. Proceeding in that order produces a different solution than those of Sections 2.3, 2.4, and 2.5. The Appendix 2 shows there is a more general solution, solving (26) and (27) for $\chi_p$, and by choice of coefficients that eliminate terms, $\chi_p = \psi_p{}^2$, equivalent to the square of (45), (46), and to (47).

Returning to Equation (39) for $p{=}p_r{\neq}0$, again, where there is only a radial momentum dependence, one finds,

$$S_{p\psi}(r) = d_2\left(\frac{\cos\left(\sqrt{2}pr + d_1\right)}{\sqrt{2}pr}\right)^{1/2} \tag{45}$$

$$\varphi_{p\psi}(x_o) = d_4\left(\cos\left(\sqrt{2}p\left(d_3 + x_o\right)\right)\right)^{1/2} \tag{46}$$

where the subscript $r$ on $p_r$ has been dropped. Note in that the limit of the solutions (45) and (46) does not equal the solution of the limit (40) and (41).

The factor $pr$ is only a scalar product, not a dot product between vectors. When squared, the solutions (45) and (46) are that of a localized spherical standing wave, not a traveling wave. This field on its own would not seem to represent a constant like $\hbar$ very well, since (45) can be either imaginary, or when squared, negative, and in either case, falling to zero at $r{=}\infty$. However, the solution of (45) and (46) is of interest, because it is unusual, and also readily quantizable.

The profile of the field $\underline{S}_{o\psi}$ from (40) is shown in Figure 1a for various values of the constants, and the profile of $\underline{S}_{p\psi}{}^2$ from (45) is shown in Figure 1b.

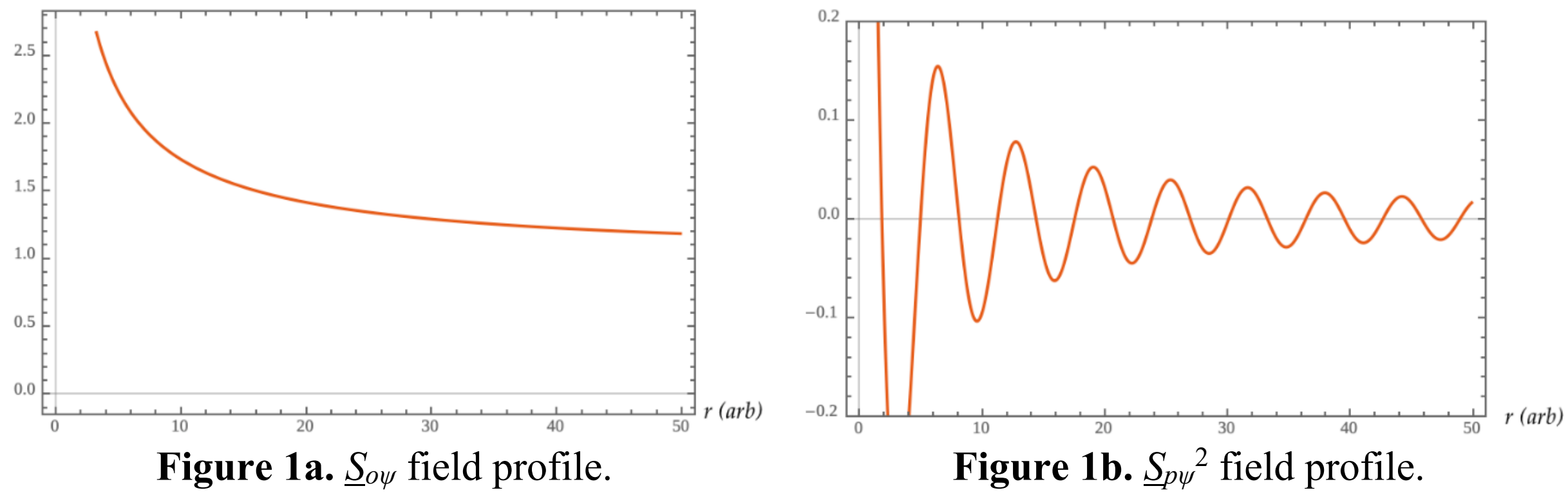

**Figure 1a.** $\underline{S}_{o\psi}$ field profile.  **Figure 1b.** $\underline{S}_{p\psi}{}^2$ field profile.

### 2.7 Canonical Quantization of the Standing Wave $\psi^2$ Solution

What may the $p=p_r\neq 0$ solutions, Equations (45-46), correspond to in a quantum field theory? The endeavor will be to attempt their canonical quantization, following the unusual steps outlined in the last section. Observe that the standing wave field $\underline{\psi_p}=\underline{S}_{p\psi}\varphi_{p\psi}$ is complex, but $\underline{\psi_p}^2$ is real, spatially oscillating and decaying with an envelope $1/r$, and also oscillating with an overall amplitude in time,

$$\psi_p^2 = \left(d_2 d_4\right)^2 \left( \frac{\cos\left(\sqrt{2}pr + d_1\right)}{\sqrt{2}pr} \right) \left( \cos\left( \sqrt{2}p\left(d_3 + x_o\right)\right)\right) \tag{47}$$

Expressing (47) in exponential form, and setting $d_1$ and $d_3$ to zero,

$$\psi_p^2 = \left(\frac{1}{2}\right)^2 \left(d_2 d_4\right)^2 \frac{1}{\sqrt{2}pr} \left( e^{i\sqrt{2}pr} + e^{-i\sqrt{2}pr} \right)\left( e^{-i\sqrt{2}px_o} + e^{i\sqrt{2}px_o} \right) \tag{48}$$

In order to make the transition from (48) to an expansion, the expansion coefficients will be associated only with the time ($x_o$) components, and formulated so as to obey a reality condition,

$$\psi_p^2 = \left(\frac{1}{2}\right)^2 \left(d_2 d_4\right)^2 \frac{1}{\sqrt{2}pr} \left( e^{i\sqrt{2}pr} + e^{-i\sqrt{2}pr} \right)\left( \tilde{a}_p e^{-i\sqrt{2}px_o} + \tilde{a}_p^* e^{i\sqrt{2}px_o} \right) \tag{49}$$

An integration over wavenumber is performed to form the complete solution as a superposition. Note that the wavenumber integral is one dimensional, since the momentum is only radial.

$$\psi^2 = \left(\frac{1}{2}\right)^2 \left(d_2 d_4\right)^2 \int \frac{dp}{(2\pi)} \frac{1}{\sqrt{2}pr} \underbrace{\left( e^{i\sqrt{2}pr} + e^{-i\sqrt{2}pr} \right)}_{2\cos(\sqrt{2}pr)} \left( \tilde{a}_p e^{-i\sqrt{2}px_o} + \tilde{a}_p^* e^{i\sqrt{2}px_o} \right) \tag{50}$$

As for the derivatives,

$$\partial_{x_o}\psi^2 = \left(\frac{1}{2}\right)^2 \left(d_2 d_4\right)^2 \int \frac{dp}{(2\pi)} (-i) \frac{\sqrt{\omega_p / c}}{r} \underbrace{\left( e^{i\sqrt{2}pr} + e^{-i\sqrt{2}pr} \right)}_{2\cos(\sqrt{2}pr)} \left( \tilde{a}_p e^{-i\sqrt{2}px_o} - \tilde{a}_p^* e^{i\sqrt{2}px_o} \right) \tag{51}$$

$$\nabla\psi^2 = \left(\frac{1}{2}\right)^2 \left(d_2 d_4\right)^2 \int \frac{dp}{(2\pi)} (i) \frac{\sqrt{\omega_p / c}}{r} \underbrace{\left( e^{i\sqrt{2}pr} - e^{-i\sqrt{2}pr} \right)}_{2i\sin(\sqrt{2}pr)} \left( \tilde{a}_p e^{-i\sqrt{2}px_o} + \tilde{a}_p^* e^{i\sqrt{2}px_o} \right) \tag{52}$$

Collecting factors carefully, and on quantizing, the fields become,

$$\psi^2 = \left(\frac{1}{2}\right) \left(d_2 d_4\right)^2 \int \frac{dp}{(2\pi)} \frac{1}{\sqrt{2\cdot 8\omega_p / cr}} \left( a_p + a_p^\dagger \right) \cos(\sqrt{2}pr) \tag{53}$$

$$\partial_{x_o}\psi^2 = \left(\frac{1}{2}\right)\left(d_2 d_4\right)^2 \int \frac{dp}{(2\pi)}(-i)\frac{\sqrt{\omega_p / 8c}}{r}\left(a_p - a_p^\dagger\right)\cos(\sqrt{2}pr) \tag{54}$$

$$\nabla\psi^2 = \left(\frac{1}{2}\right)\left(d_2 d_4\right)^2 \int \frac{dp}{(2\pi)}(-1)\frac{\sqrt{\omega_p / 8c}}{r}\left(a_p + a_p^\dagger\right)\sin(\sqrt{2}pr) \tag{55}$$

The Hamiltonian density is,

$$\mathcal{H} = \frac{1}{8}\beta^2\left(\partial_{x_o}\psi^2\right)^2 + \frac{1}{8}\beta_{\psi\psi}^2\left(\nabla\psi^2\right)^2 \tag{56}$$

and the total energy,

$$E = \frac{1}{8^2}\beta^2\left(d_2 d_4\right)^4\left(\frac{1}{2}\right)^2 \int d^3r \frac{dpdk}{(2\pi)^2}\frac{\sqrt{\omega_p\omega_k}}{cr^2}\times \\ \begin{pmatrix} (-1)\cos(\sqrt{2}pr)\cos(\sqrt{2}kr)\left(a_p - a_p^\dagger\right)\left(a_k - a_k^\dagger\right) \\ + \\ (+1)\sin(\sqrt{2}pr)\sin(\sqrt{2}kr)\left(a_p + a_p^\dagger\right)\left(a_k + a_k^\dagger\right) \end{pmatrix} \tag{57}$$

Multiplying out (57), rearranging terms, integrating over spatial angles, and cancelling the $r^2$ from the denominator stemming from the spatial integral,

$$E = \frac{\pi}{8^2}\beta^2\left(d_2 d_4\right)^4 \int \frac{dpdk}{(2\pi)^2}\frac{\sqrt{\omega_p\omega_k}}{c}\times \\ \begin{pmatrix} \left(a_p a_k^\dagger + a_p^\dagger a_k\right)\underbrace{\int dr\left(\sin(\sqrt{2}pr)\sin(\sqrt{2}kr) + \cos(\sqrt{2}pr)\cos(\sqrt{2}kr)\right)}_{(2\pi)\delta(k-p)} \\ + \\ \left(a_p a_k + a_p^\dagger a_k^\dagger\right)\underbrace{\int dr\left(\sin(\sqrt{2}pr)\sin(\sqrt{2}kr) - \cos(\sqrt{2}pr)\cos(\sqrt{2}kr)\right)}_{-(2\pi)\delta(k+p)} \end{pmatrix} \tag{58}$$

The interpretation of the formally non-convergent integral is shown in the under-brackets of (58). Continuing,

$$E = \frac{\pi}{8^2}\beta^2\left(d_2 d_4\right)^4 \int \frac{dp}{(2\pi)}\frac{\omega_p}{c}\left(\left(a_p a_p^\dagger + a_p^\dagger a_p\right) - \underbrace{\left(a_p a_{-p} + a_p^\dagger a_{-p}^\dagger\right)}_{0}\right) \tag{59}$$

$$E=\frac{\pi}{32}\beta^2\left(d_2d_4\right)^4\int\frac{dp}{(2\pi)}\frac{\omega_p}{c}\left(a_p^\dagger a_p+\underbrace{\frac{(2\pi)}{2}\delta^{(1)}(0)}_{\infty}\right) \tag{60}$$

$$:E:=\frac{\pi}{32}\beta^2\left(d_2d_4\right)^4\int\frac{dp}{(2\pi)}\frac{\omega_p}{c}a_p^\dagger a_p \tag{61}$$

Well-defined quantization immediately follows if the integrals are restricted to range from $p$=0 to $p$=+∞, where equivalently the creation and annihilation operators for states with subscripts that are negative are taken to be zero in (59), so (60) results. Normal ordering eliminates the vacuum term in (60) giving the field quanta and their energies, (61).

One may take this as a condition necessary to achieve quantization with no further explanation. However, a physical rationale will be offered, below.

### 2.7.1 A Cosmological Interpretation

Standing waves are formed from oppositely directed traveling waves. The classical solution (47) does not rely on the $-p$ solutions to combine with the $+p$ solutions to form the standing wave – it is a steady state spherical standing wave in and of itself already, describing the situation long after transient behavior is over, after setup. Now visualize the early transient behavior whereby the standing wave comes about. The acton emits a traveling spherical wave from its origin that propagates outward into space, and to develop the spherical standing wave (47), this emitted wave would have to somehow reflect off of a perfectly symmetrical spherical boundary and be directed back to the origin to interfere with the emitted wave. However, there is no physical boundary to provide reflection, so how can this occur? Assume closed curvature in the Universe during the initial field set-up, before Inflation drives the Universe to the critical density, and the associated flat spacetime. Impose periodic boundary conditions with allowed wavevectors incremented by $2\pi/L$, where $L$ is the period of the boundary taken to the limit $L$=∞. Each point on the traveling spherical wavefront emitted from the origin at $r$=0 travels full circle, returning from $r$=∞ to the origin from the *opposite* side it was emitted from, and then interferes with the emitted wave. So, for the $+p$ states, which are initially outgoing waves emitted from the origin,

$$\underbrace{\frac{e^{+i\sqrt{2}|p|r-i\sqrt{2}\omega_p t}}{2^{3/2}\,|\,p\,|\,r}}_{emitted,\,r=0}+\underbrace{\frac{e^{-i\sqrt{2}|p|r-i\sqrt{2}\omega_p t}}{2^{3/2}\,|\,p\,|\,r}}_{returning,\,r=\infty}\propto\frac{\cos(\sqrt{2}\,|\,p\,|\,r)}{\sqrt{2}\,|\,p\,|\,r}\cos(\sqrt{2}\omega_p t) \tag{62}$$

and (39) is becomes the standing spherical wave (47) for $+p$ after taking the real part.
The $-p$ states must then be initially inbound waves emitted from a spherical boundary at $r$=∞ traveling to the origin at $r$=0, and forming the standing wave with the wave also traveling inbound from infinity that passed through the origin from the opposite side,

$$\underbrace{\frac{e^{-i\sqrt{2}|p|r-i\sqrt{2}\omega_p t}}{-2^{3/2}\,|\,p\,|\,r}}_{emitted,r=\infty}+\underbrace{\frac{e^{i\sqrt{2}|p|r-i\sqrt{2}\omega_p t}}{-2^{3/2}\,|\,p\,|\,r}}_{passthrough,r=0}\propto\frac{\cos(\sqrt{2}\,|\,p\,|\,r)}{-\sqrt{2}\,|\,p\,|\,r}\cos(\sqrt{2}\omega_p t) \tag{63}$$

So, it can be seen that the real part of (63) is the solution (47) for $-p$.

However, the case of a distant boundary of sources miraculously located for simultaneous convergence on the acton center feels as unphysical as a distant reflective boundary, even though mathematically the solution is allowed, and the steady state behavior differs by only a phase of $\pi$ between $+p$ and $-p$.

If there had in the distant past been closed curvature in the Universe, in the equilibration period after the Big Bang, but before Inflation at $10^{-36}$ s, the acton field could have been set up, interfered with itself, and then expanded during Inflation.

What is the acton origin? The following is postulated:

i. In all the above scenarios applied to cosmology, the acton origin would be every point in the universe, if the Cosmological Principle is to be obeyed. The Planck's constant experienced on the cosmological scale would be then be a spatial average, something like the root-mean square value of (45) and (40), subject to boundary conditions that prevent its divergence, leaving a cosmological time dependence per (46) and (41). This is the background value of $\hbar$. Variations on this idea of averaging is treated in Appendix 3, Section 3.5.2, and also reference [43].

ii. The infinite number of cosmological origins (every point in space) may also be thought of as coinciding with a location of *initially* infinite energy density that *does not persist*.

iii. The field at a single origin in isolation is divergent per (47). However, the average of all origins includes an infinite number of effectively zero values, and the field can be quantized without divergence. Therefore, the spatial averaging produces a finite value with only a time dependence, see Appendix 3.

iv. On astrophysical scales, in this epoch, where localized energy densities are not uniform, that *do persist*, such as supermassive black holes, neutron stars, and white dwarfs, then (40-41) and (45-46) would represent more local perturbations those objects around the background value, so not subject to an averaging process.

Whatever the acton origin is, to quantize this field, it must be thought of as the source of the activity, the integrals $dp$ range from $p$=0 to ∞, the operators with negative subscripts are zero, and the second term of (59) is zero.

### 2.7.2 A Comment on Coupling to Gravity and Closed Curvature

Why has this author not started with the case in which the field being investigated is coupled to gravity, either minimally via the metric, or non-minimally, directly to the Einstein-Hilbert action. The spacetime in these derivations is flat. This is certainly a valid criticism.

The reason is that what is being examined is the most tractable possible case, as finding solutions coupled to gravity becomes orders more difficult, having to find solutions to the Einstein Field Equations and the acton field simultaneously. That may very well be the problem that has to be solved, ultimately, and later. It is much easier investigate the situation under the assumption that the curvature of the universe is “sufficiently flat” per Section 2.9 (though not exactly flat, per Section 2.7.1), and develop a base set of solutions, for which the coupled solutions with gravity may eventually be compared. There are also an infinite number of ways the acton field may be coupled to gravity, and it is not clear at this time which couplings are the most appropriate to investigate, although minimal coupling would be a reasonable one to start with, and only because it is the second simplest. If the author started the work there, the criticism would then concern justifying the reason for that specific case, but also, there already exist many solutions for scalar fields coupled to gravity in many ways.

In addition, this author’s primary interest is in investigating how a scalar field, representing a constant, “supports” itself, and other fields, the simplest “other field” being another scalar field.

However, an interpretation has been given here, that during a period of closed curvature, the outgoing waves complete their full journey, become incoming waves, and interfere with those still outgoing, followed by inflation and continued expansion. In that epoch, undoubtedly, the interaction with gravity may be very strong. In the present epoch, though now very close to perfect flatness, perhaps not exactly, this process must still be happening. This makes the near-flat spacetime case a worthy starting point, and the approximations used here worth considering.

### 2.7.3 Mathematical Details for Completion of Section 2.7

The commutation relationships used were,

$$\left[a_p, a_k^{\dagger}\right] = (2\pi)\delta^{(1)}(p-k) \tag{64}$$

$$\left[a_p^{\dagger}, a_k^{\dagger}\right] = \left[a_p, a_k\right] = 0 \tag{65}$$

from which may be derived the relationship between conjugate variables,

$$\left[\psi^2(r), \psi^2(r')\right] = \left[\partial_{x_o}\psi^2(r), \partial_{x_o}\psi^2(r')\right] = 0 \tag{66}$$

$$\left[\psi^2(r), \partial_{x_o}\psi^2(r')\right] = i\left(\frac{\left(d_2 d_4\right)^2}{16\sqrt{2}}\right)\delta^{(1)}(r-r') \tag{67}$$

where (65) is arrived at in interpreting the following integral that results during the analysis,

$$\int dp \frac{\cos\left(\sqrt{2}pr\right)\cos\left(\sqrt{2}pr'\right)}{rr'} = (2\pi)\delta^{(1)}(r-r') \tag{68}$$

The final result (61) resembles those of normal free fields, save for the pre-factors, and that the operators are left dimensioned.

### 2.8 Calibration of Coefficients – A Demonstration

This section deals with an example of the inputs necessary to calibrate a model. It is conjectured here that the static, non-propagating modes that couple to other fields in the derivative terms is, from (41) or (38), the square of (69). This is because it is the only solution for $\hbar$ that is positive and real at all times. The free-field solution, and the standing-wave solutions result during temporary adjustments to this non-propagating mode.

From (69) follow other relations to be used in the calibration of the coefficients,

$$\underline{\hbar}_{o\psi}(r,t) = \beta\sqrt{b_1 b_3\left(1+\frac{b_2}{b_1}x_o\right)\left(1+\frac{b_4/b_3}{r}\right)} \tag{69}$$

$$\underline{\hbar}_{o\psi}(\infty,t) = \beta\sqrt{b_1 b_3\left(1+\frac{b_2}{b_1}ct\right)} \tag{70}$$

$$\hbar_m = \underline{\hbar}_{o\psi}(\infty,t_H) \tag{71}$$

$$\frac{\partial_t \underline{\hbar}_{o\psi}(r,t)}{\underline{\hbar}_{o\psi}(r,t)} = \frac{1}{2}\frac{c}{\left(b_1/b_2+ct\right)}\Bigg|_{t_H} = -\frac{\partial_t \alpha}{\alpha} = 4.73\times10^{-18}[yr^{-1}] \tag{72}$$

The value that is measured now is $\hbar_m$ and $t_H$ is the Hubble time. The position dependence of $\underline{\hbar}_{o\psi}$ is thought provoking, see [25] and its relevant equation (97) reproduced in Section 4.0. Equation (122) describes the variation of scalar fields that (may represent a constant) coupled to a gravitational field when gravity is weak, and [27] is an example, wherein other references are given on the matter. Equation (123) is derived in this paper from (69), although gravity was only obliquely involved, in the curvature discussion, Section 2.7.1. They are very similar in appearance.

In (72), $\alpha$ is the fine structure constant, and its variation is constrained to ~ $10^{-17}$ per year [4-6]. The value chosen for use in (72) is on that order, but the choice of sign was made completely arbitrarily, to get order of magnitude estimates for the calibration demonstration. In one paper it is concluded that $\alpha$ is smaller in the past [4], opposite the trend chosen for this demonstration in (72). Most works only bound the possible variation about zero.

For $t$=$t_H$, one then finds when attributing all the variation in the fine structure constant to Planck's constant,

$$b_1/b_2 = 10^{33}[m] \tag{73}$$

From (70-71) and (73), again putting $t = t_H$, one finds,

$$\beta^2 b_2 b_3 = \frac{\hbar_m^2}{\left(b_1 / b_2 + ct_H\right)} \sim 1.11\times 10^{-101} [\hbar^2 / m] \tag{74}$$

In (74) and the equations to follow, when $\hbar$ appears in a square bracket, should be understood to mean "units of Planck's constant J-s".

It was conjectured in Section 2.3 and 2.4 that the energy density associated with the constant $\hbar$ when the field has no momentum is that of the vacuum, and the follow-up comes below.

Note the following:

i. Arguments where given that the solution (69) for Planck's constant is associated with vacuum energy, and it is a vacuum solution.
ii. Planck's constant is small.
iii. The Cosmological Constant, the gravitational vacuum energy density, and the only vacuum for which an experimental value of an energy density exists, is also small.
iv. Quantum mechanical effects are thought to be critically important at the energy scale of the pre-inflation universe, and at the singularities of black holes.
v. Planck's constant is involved in all of the known quantum mechanical interactions, and also sets the magnitudes of a system of units for quantum gravity theories.
vi. The non-propagating solutions are the most reasonable candidate fields to assign to the present epoch. Fundamental constants must be changing very slowly. Therefore, there are a minimal number of particles needed to readjust them at this time, so we do not routinely just find them.
vii. As originally introduced by Einstein, the Cosmological Constant does not change in value as the universe expands (though in quintessence models, *k*-essence, and others, it does).
viii. The electromagnetic vacuum energy density is formally divergent, and $10^{120}$ orders of magnitude larger than the cosmological constant when an upper bound of the Planck frequency is imposed. The vacuum catastrophe is still unresolved, and is the premier example concerning the viability of field theory, as its gravitational influence should be enormous, but it is not sensed at all. This is indicative of the lack of some fundamental understanding of fields, so is also avoided here.

Therefore, from (37), very far from the acton origin, the present value of the gravitational Cosmological Constant is this authors choice to equate to the energy density of the field $\hbar^2 = \beta^2\chi$,

$$\underline{H}_{o\chi} \Big|_0^0 (r = \infty) = \frac{\beta^2 (b_2 b_3)^2}{8} = \Lambda = 5.63\times 10^{-10} [J / m^3] \tag{75}$$

From (74) and (75),

$$b_2 b_3 = 3.86\times10^{92}[J^{-1}m^{-2}s^{-2}] \tag{76}$$

$$\beta = 1.70\times10^{-97}[kg^{3/2}m^{7/2}/s] \tag{77}$$

$$b_1 b_3 = 3.86\times10^{125}[J^{-1}m^{-1}s^{-2}] \tag{78}$$

From (69), and independent of the time-dependence choice of the fine structure constant,

$$\frac{\partial_r \underline{\hbar}_{o\psi}}{\underline{\hbar}_{o\psi}} = -\frac{1}{2}\frac{b_4}{b_3}\frac{1}{r^2}\frac{1}{\left(1+\frac{b_4/b_3}{r}\right)}\Bigg|_{R_o} \approx \frac{\delta\hbar}{\hbar}\frac{1}{\Delta R_o} = -3.5\times10^{-15}[m^{-1}] \tag{79}$$

$$b_4/b_3 = 1.62\times10^{8}[m] \tag{80}$$

The numerical value in (79) comes from the work of Hutchin [19] where it was surmised $\hbar$ varies by 21 ppm across the Earth's orbit of radius $R_o \sim 1.52\times10^{11}$ m, and the difference between the maximum and minimum radii $\Delta R_o = 6\times10^9$ m. One might have used the bounding values of [26,27,29] for the example, but those pertain to the fine structure constant, and also, some appear to be null results, or not to be statistically significant.

The author mentions that recent work on spectral analysis of the white dwarf G191-B2B suggests that the fine structure constant increases slightly in strong gravitational fields [43]. This is the opposite of what would be expected if Planck's constant were solely responsible.

From (76), (78) and (80), Planck's constant is positive, and decreases with distance from the origin using the data of [19].

### 2.9 Coupling to Other Fields

The equations of motion for the fields coupled by the dynamical terms can only be solved analytically, with some approximations. The Lagrangian will be written in terms of the "supporting field" $\chi$, and the "supported field" of the Klein-Gordon type $\varphi$, so-called because $\chi$ supports the dynamical terms of $\varphi$, and also, because these fields did not get along very well, and $\chi$ had higher income.

Consider (16) and massless fields, with an action (81a). $\chi$ is non-minimally coupled to gravity via some function $f(\chi)$, but $\varphi$ is minimally coupled, and $q$ is a coupling constant.

$$S = \int \frac{1}{2\kappa} q f(\chi) R + \frac{1}{8}\beta^2 \chi g^{\mu\nu}\nabla_\mu\chi\nabla_\nu\chi + \frac{1}{2}\beta^2\chi g^{\mu\nu}\nabla_\mu\varphi\nabla_\nu\varphi + \mathcal{L}_m\{\chi\}\sqrt{-g}\, d^4\mathrm{x} \tag{81a}$$

When gravity is weak and matter is dilute, the "sufficiently flat" approximation is that spacetime is very nearly flat, $\chi$ is still weakly affected by matter and the curvature in some manner, but $\varphi$ is affected only through the coupling to $\chi$. Thus, $\chi$ can be treated as though it changes independently, and one may work with the Lagrangian (81b) to estimate the impact on $\varphi$,

$$\mathcal{L} = \frac{1}{2}\beta^2 \chi \partial_u \varphi \partial^u \varphi + \frac{1}{8}\beta^2 \partial_u \chi \partial^u \chi \tag{81b}$$

The equations of motion for $\chi$ and $\varphi$ are, respectively,

$$\frac{1}{4}\left(\ddot{\chi} - \nabla^2 \chi\right) - \frac{1}{2}\left(\dot{\varphi}^2 - \left(\nabla \varphi\right)^2\right) = 0 \tag{82}$$

$$\left(\dot{\chi}\dot{\varphi} + \chi\ddot{\varphi}\right) - \left(\nabla\chi \cdot \nabla\varphi + \chi\nabla^2\varphi\right) = 0 \tag{83}$$

The variable Planck's constant field will now be shown to cause an alteration in the solution of the supported field from its free state.

The situations when the second term of (82) is zero are the easiest to solve for. Then $\varphi$ is close to the form of a plane wave, $\varphi$ does not influence the form of $\chi$, and solutions for the latter already developed may be used, namely, those of (35) and (36). Therefore, the equation of motion (83) can be solved with $\chi$ as an input from its solution, in isolation, for the limit $p=0$, taken first.

The following will be needed,

$$\chi = \underline{\chi}\Big|_0^0 \tag{84}$$

$$\underline{\chi}\Big|_0^0 = \left(b_1 + b_2 x_o\right)\left(b_3 + \frac{b_4}{r}\right) \tag{85}$$

$$\underline{\dot{\chi}}\Big|_0^0 = b_2\left(b_3 + \frac{b_4}{r}\right) \tag{86}$$

$$\nabla\, \underline{\chi}\Big|_0^0 = -b_4 \frac{\left(b_1 + b_2 x_o\right)}{r^2} \tag{87}$$

$$\frac{\underline{\dot{\chi}}\Big|_0^0}{\underline{\chi}\Big|_0^0} = \frac{b_2}{\left(b_1 + b_2 x_o\right)} \tag{88}$$

$$\frac{\nabla\, \underline{\chi}\Big|_0^0}{\underline{\chi}\Big|_0^0} = \frac{-b_4}{\left(b_3 r^2 + b_4 r\right)} \tag{89}$$

$$\hbar^2 = \beta^2 \chi \tag{90}$$

For $r=\infty$, the gradient of $\chi$ is zero, and (83) becomes,

$$\frac{\dot{\chi}}{\chi}\dot{\varphi} + \ddot{\varphi} - \nabla^2 \varphi = 0 \tag{91}$$

For very early times $x_o=0$, and from (89), (91) becomes, using the calibration demonstration values for $b_2/b_1$, writing $\varphi=T(t)Z(r)$, and separating variables,

$$\omega_h = \frac{b_2}{b_1} c \sim 10^{-25} [s^{-1}] \tag{92}$$

$$\omega_h \partial_t T + \partial_t^2 T = -\omega_p^2 T \tag{93}$$

$$\nabla^2 Z = -p^2 Z \tag{94}$$

$$\varphi = \left( A e^{i\bar{p}\cdot\bar{r}} + B e^{-i\bar{p}\cdot\bar{r}} \right) \left( C e^{-\frac{t}{2}\left(\sqrt{\omega_h^2 - 4\omega_p^2} + \omega_h\right)} + D e^{\frac{t}{2}\left(\sqrt{\omega_h^2 - 4\omega_p^2} - \omega_h\right)} \right) \tag{95}$$

One sees that the supported field $\varphi$ has been altered from the free-field solution due to the variable Planck's constant. For $\omega_h=0$, the supported field $\varphi$ would be a pure planewave and therefore a free field.

Based on the calibration of the constants, even at a time of $t_H \sim 10^{17}$ [s], the solution (95) is almost that of a plane wave for any reasonable value of frequency $\omega_p$, but with a very small decay in time, controlled by $\exp(-\omega_h t/2)$, the parameter $\omega_h$ describing the temporal variation of Planck's constant.

One finds with the calibrated parameters, that the increase/decrease of $\hbar$ with time causes the supported field amplitudes to decay/grow in time. The signs and trend here are dictated by the arbitrarily chosen sign in (72).

The plane wave condition for the second term of (82) being approximately zero is therefore satisfied, and for $\omega_h t$ so small, it is clear how to quantize (95) for early times – neglect $\omega_h t$ and proceed with canonical quantization.

## 3.0 Path Integral for a Position Dependent Planck's Constant

A related subject is the path integral, allowing one to examine the impact of a position-dependent Planck's constant on classical trajectories, and to compare to some of the results from the dynamical field study.

A derivation is shown for how the integrand in the path integral exponent becomes $L_c/\hbar(r)$, where $L_c$ is the classical action. The path that makes stationary the integral in the exponent is termed the "dominant" path, and deviates from the classical path systematically due to the position dependence of $\hbar$. The changes resulting in the Euler-Lagrange equation, Newton's first and second laws, Newtonian gravity, the Friedmann equation with a Cosmological Constant, and the impact on gravitational radiation for the dominant path are shown and discussed. Total frequency, not energy, is conserved, explained below.

### 3.1 Derivation of an Alternative Path Integral

The development to follow will depend on some prior results found in [25]. Equation (96) is the anticommutator-symmetrized Hermitian frequency-conserving operator controlling unitary time-evolution,

$$\hat{F}_h = -\frac{1}{2m}\frac{1}{2}\left\{\hbar(\overline{r}),\nabla^2\right\} + \frac{V(\overline{r})}{\hbar(\overline{r})} = \frac{\hat{p}_\hbar^2}{2m} + V_\hbar(\overline{r}) \tag{96}$$

and the analog of momentum with units of [kg-m/s/$\hbar^{1/2}$] is,

$$\hat{p}_\hbar = \frac{1}{i}\sqrt{\frac{1}{2}\left\{\hbar(\overline{r}),\nabla^2\right\}} \tag{97}$$

In this approach, total frequency is conserved, not energy. Equations (96-97) were derived for the condition that $\hbar$ had no explicit time dependence. The completeness operator in the position representation that will result in summation over every possible path at each time-slice is,

$$\int dx_j(t_j)\,|\,x_j,t_j\rangle\langle x_j,t_j\,| = 1 \tag{98}$$

Using (98) by repeated insertion $N$ times (for $N$ time slices) between the brackets of the transition amplitude for a particle initially at $(x_i, t_i)$ to be found at $(x_f, t_f)$, one may write for the amplitude,

$$\langle x_f,t_f\,|\,x_i,t_i\rangle = \int\prod_{j=1}^{N-1} dx_j(t_j)\prod_{k=0}^{N-1}\langle x_{k+1},t_{k+1}\,|\,x_k,t_k\rangle \tag{99}$$

The completeness operator will be needed for the equivalent of the momentum representation to be used,

$$\int dp_\hbar\,|\,p_\hbar\rangle\langle p_\hbar\,| = 1 \tag{100}$$

A general amplitude in (99) will now be examined. Using the time evolution operator followed by approximation to first order,

$$\langle x_1, t_o + \Delta t \,|\, x_o, t_o \rangle = \langle x_1 \,|\, \exp\left(-i\hat{F}_\hbar \Delta t\right) |\, x_o \rangle$$
$$\approx \langle x_1 \,|\, \exp\left(-i\frac{\hat{p}_\hbar^2}{2m}\Delta t\right)\exp\left(-iV_\hbar(\hat{x})\Delta t\right) |\, x_o \rangle \tag{101}$$

Inserting (100) into (101) and acting with the operators,

$$\langle x_1, t_o + \Delta t \,|\, x_o, t_o \rangle = \int dp_\hbar \langle x_1 \,|\, \exp\left(-i\frac{\hat{p}_\hbar^2}{2m}\Delta t\right) |\, p_\hbar \rangle \langle p_\hbar \,|\, \exp\left(-i\frac{V(\hat{x})}{\hbar(\hat{x})}\Delta t\right) |\, x_o \rangle$$
$$= \int dp_\hbar \langle x_1 \,|\, p_\hbar \rangle \langle p_\hbar \,|\, x_o \rangle \exp\left(-i\frac{p_\hbar^2}{2m}\Delta t\right)\exp\left(-i\frac{V(x_o)}{\hbar(x_o)}\Delta t\right) \tag{102}$$

The two needed $p_\hbar$ eigenfunctions, with factors of $\hbar^{1/2}$ dividing the exponent to produce the right units for the approximate basis become,

$$\langle p_\hbar \,|\, x_o \rangle \approx \frac{\exp\left(-ip_\hbar x_o / \sqrt{\hbar(x_o)}\right)}{\sqrt{2\pi\hbar(x_o)}}$$
$$\langle x_1 \,|\, p_\hbar \rangle \approx \frac{\exp\left(ip_\hbar x_1 / \sqrt{\hbar(x_1)}\right)}{\sqrt{2\pi\hbar(x_1)}} \approx \frac{\exp\left(ip_\hbar x_1 / \sqrt{\hbar(x_o)}\right)}{\sqrt{2\pi\hbar(x_o)}} \tag{103a-b}$$

The approximate basis above is justified in [25] where it was shown that for a mild enough gradient in $\hbar$, the free particle wavefunctions are approximately planewaves, and the Ehrenfest theorem relating the position expectation value time derivative to the momentum is exactly retained. The approximation will break down if the gradient becomes too large. Substituting (103a-b) into (101) and (102) and integrating,

$$\langle x_1, t + \Delta t \,|\, x_o, t_o \rangle = \int \frac{dp_\hbar}{2\pi\hbar(x_o)} \exp\left(-i\left\{\frac{p_\hbar(x_o - x_1)}{\sqrt{\hbar(x_o)}} + \left(\frac{p_\hbar^2}{2m} + \frac{V(x_o)}{\hbar(x_o)}\right)\Delta t\right\}\right)$$
$$= \frac{1}{2\hbar(x_o)}\sqrt{\frac{m}{2\pi i \Delta t}} \exp\left(\frac{i}{\hbar(x_o)}\left\{\frac{1}{2}m\underbrace{\left(\frac{x_1 - x_o}{\Delta t}\right)^2}_{\dot{x}_o^2} - V(x_o)\right\}\Delta t\right) \tag{104}$$

The term in the curly brackets of (104) is the classical Lagrangian $L_c$, only now divided by the position dependent $\hbar$,

$$\langle x_1, t_1 \,|\, x_o, t_o \rangle = \frac{A}{\hbar(x_o)} \exp\left(i\frac{L_c(x_o)}{\hbar(x_o)}\Delta t\right) \tag{105}$$

Now substituting (105) into (99), taking the limit $N\rightarrow\infty$, and passing the resulting sum in the exponent to an integral, the new form of the path integral is,

$$\langle x_f, t_f \mid x_i, t_i \rangle = \int D_\hbar x(t) \exp\left( i \int \underbrace{\frac{dt}{\hbar(x(t))}}_{d\tau(x)/\hbar_\infty} L_c(x(t), \dot{x}(t)) \right)$$

$$D_\hbar x(t) = \lim_{N\to\infty} \prod_{j=1}^{N-1} A \frac{dx(t_j)}{\hbar(x(t_j))} \qquad \text{(106a-b)}$$

Normally, the $\hbar$ that appears in the denominator of the exponent is a constant, but in (106a) it is not, and is being integrated. The complication of the product $1/\hbar(x_j)$ in the pre-factor of (106b) appears in $D_\hbar x(t)$, however, this is not important in what follows in Section 3.1.1, and after.

### 3.1.1 Meaning of a Variable Planck's Constant

Noting (105), equation (106a) and (106b) have been written to provide an interpretation for what a variation in Planck's constant actually means in this context – it is related to variations in the rate of time passage over the classical path, which must be made stationary in combination with the Lagrangian as part of the action. Unusual forces not normally present in classical physics result, to be discussed.

Unlike relativity, here, an observer at rest in the field of $\hbar$ with respect to another in motion through it do not measure different rates of time passage, and both observe the same altered trajectory. The amplitude of the transition from an initial state at $x_i$ to the final state at $x_f$ behaves as if there is an effective time increment between, depending on the value of $\hbar(x_i)$.

The expressions (105) and (106a-b) were not derived by straightforward substitution of a position dependent Planck's constant into the path integral, and to do so would be rather trivial. Here, a different form of the pre-factor results (106b). The result shown came by way of the considerations that lead to Equations (96) and (97), in a non-field-theoretic approach. There are not two (or more) coupled fields sharing the total energy, so the approach is non-traditional. The rationale for this, was that what constants actually are is not understood, and so non-traditional approaches are worthy of investigation and memorialization. Here, and in [25], the non-traditional approach will be the conservation of total frequency, as opposed to energy.

### 3.2 Euler-Lagrange Equation for Dominant Path and Total Frequency Conservation

The usual argument is to say that the classical path is the one that makes the classical action in the exponent of the path integral stationary, all other paths cancelling by rapid oscillations in the limit that the constant $\hbar\rightarrow 0$. That will still be the case if the entire function $\hbar$ in (106a) goes to zero.

To be more precise, the classical trajectory is recovered from the path integral when the classical action is much larger than the constant $\hbar$, due to mass or energies becoming large. If $\hbar$ had no positional dependence, the classical path would be found by making stationary the Lagrangian in the exponential term in (106a).

The “dominant path” will be used to refer to one that makes the integral in the exponent of (106a) stationary when $\hbar$ varies. The trajectories around the dominant path are systematically different from the true classical path, due to the variation of $\hbar$.

The condition for a stationary exponent in (106a) in terms of generalized coordinates is given by a modified form of the Euler-Lagrange equation,

$$L = L(\underbrace{q_1(t), q_2(t), q_3(t)}_{q_i}, \underbrace{\dot{q}_1(t), \dot{q}_2(t), \dot{q}_3(t)}_{\dot{q}_i})$$

$$\hbar = \hbar(q_1(t), q_2(t), q_3(t))$$

$$\frac{d}{dt}\frac{\partial\left(L_c/\hbar\right)}{\partial \dot{q}_i} - \frac{\partial\left(L_c/\hbar\right)}{\partial q_i} = 0$$

$$\frac{d}{dt}\frac{\partial(L_c/\hbar)}{\partial \dot{q}_i} = \frac{d}{dt}\left(\overbrace{\frac{1}{\hbar}\frac{\partial L_c}{\partial \dot{q}_i} - \frac{L_c}{\hbar^2}\underbrace{\frac{\partial \hbar}{\partial \dot{q}_i}}_{0}}^{p_i}\right) = \frac{1}{\hbar}\frac{d}{dt}\frac{\partial L_c}{\partial \dot{q}_i} - \frac{1}{\hbar^2}\frac{\partial L_c}{\partial \dot{q}_i}\underbrace{\sum_j \frac{\partial \hbar}{\partial q_j}\dot{q}_j}_{\bar{\nabla}\hbar\cdot\dot{\bar{q}}}$$

$$\frac{\partial\left(L_c/\hbar\right)}{\partial q_i} = \frac{1}{\hbar}\frac{\partial L_c}{\partial q_i} - \frac{L_c}{\hbar^2}\frac{\partial \hbar}{\partial q_i}$$

(107a-f)

$$\frac{d}{dt}\frac{\partial L_c}{\partial \dot{q}_i} - \frac{\partial L_c}{\partial q_i} = \left(\bar{\nabla}\ln\hbar\cdot\dot{\bar{q}}\right)\frac{\partial L_c}{\partial \dot{q}_i} - L_c\frac{\partial \ln\hbar}{\partial q_i}$$

The left side of (107f) are the usual Euler-Lagrange terms, but the right that is usually zero is no longer. The classical equation of motion is recovered when the logarithmic derivative of $\hbar$ vanishes. In the absence of an external potential, classical conjugate momentum $p_{ic}$ and modified conjugate momentum $p_i$ are not conserved due to the position dependence of $\hbar$.

Equation (108) shows that the total frequency $W = H_c/\hbar$ is conserved and not the total classical energy. Taking the total time derivative of $L_c/\hbar$ and using (107c),

$$-\frac{\partial\left(L_c/\hbar\right)}{\partial t} = \frac{d}{dt}\left(\sum_i \dot{q}_i\frac{\partial\left(L_c/\hbar\right)}{\partial \dot{q}_i} - L_c/\hbar\right) = \frac{d}{dt}\frac{H_c}{\hbar} = \frac{dW}{dt} = 0 \qquad (108)$$

For a conserved $W$, on any trajectory in which the value of $\hbar$ is equal at the start and end, the total energy is restored, though not conserved along the trajectory.

### 3.3 Newton's First and Second Law for Dominant Path

From (107f), the equation of motion for the path that makes the Lagrangian in (106a) stationary is a modified Newton's second law,

$$m\ddot{\bar{x}} = -\bar{\nabla} V_c + m\left(\bar{\nabla}\ln\hbar\cdot\dot{\bar{x}}\right)\dot{\bar{x}} - (\bar{\nabla}\ln\hbar)\left(\frac{1}{2}m\,|\,\dot{\bar{x}}\,|^2 - V_c\right)$$

$$m\ddot{x} = F_c + \frac{\partial\ln\hbar}{\partial x}H_c \qquad \text{(109a-b)}$$

where (109a-b) are in 3-D and 1-D, respectively. $H_c=T_c+V_c$ is the classical total energy, and can be written as (109b) only in 1-D. With no external potential, the equation (109b) becomes,

$$\ddot{x} = \frac{\partial\ln\hbar}{\partial x}\left(\frac{1}{2}\dot{x}^2\right) \qquad \text{(110)}$$

From Equation (110), if the particle is at rest, it stays at rest, per the first half of Newton's first law. For the second half of Newton's first law, it is found that a particle in motion tends to accelerate or decelerate, depending on the functional form of $\hbar$. Therefore, momentum is not conserved.

As an example, assuming the logarithmic derivative of $\hbar$ is a constant $k$, with no other forces present per (110), one finds,

$$\hbar = \hbar_o e^{kx}$$

$$x(t) = -\frac{2}{k}\ln\left(\frac{c_1+kt}{c_2}\right)$$

$$\dot{x}(t) = -\frac{2}{c_1+kt} \qquad \text{(111a-d)}$$

$$\ddot{x}(t) = \frac{2k}{(c_1+kt)^2}$$

From (111c), one sees that once in motion, the particle can never be at rest unless an infinite amount of time has passed. The acceleration may increase, or decrease depending on the sign of $k$. In Figure 2, Equations (111b-c) are plotted for $k=\pm1$, and $c_1= c_2 = 1$ arbitrary unit.

An object initially moving in the direction of lower Planck's constant decelerates asymptotically to zero velocity, but never fully stops. If it is initially moving towards higher Planck's constant, it accelerates in that direction to infinite velocity, where after the position becomes undefined.

Clearly, the classical energy is no longer conserved, as there is a tendency for matter to receive an added push through space in the direction of increasing $\hbar$ at the gentlest disturbance from rest in that direction, for large $|k|$.

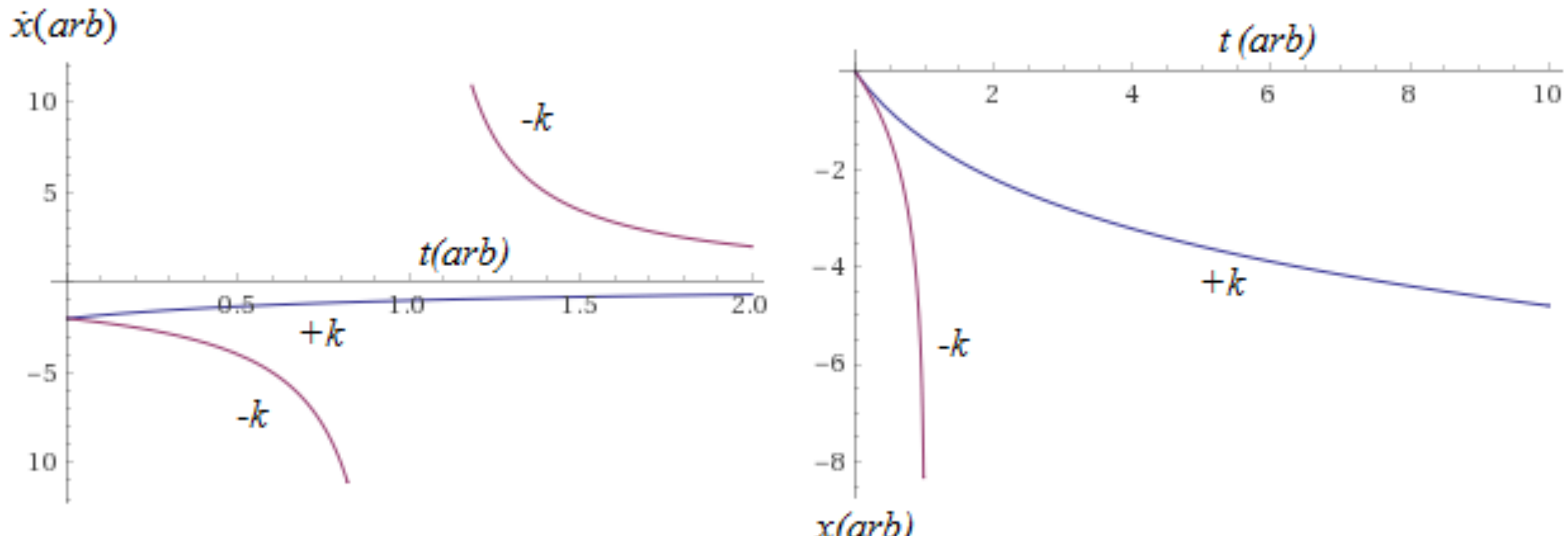

**Figure 2.** (Left) Velocity as a function of time. (Right) Position as a function of time.

### 3.4 Newtonian Gravity for Dominant Path (NGDP)

From (109a), for a mass $m$ in a gravitational potential caused by $M$,

$$\begin{aligned}
m\ddot{\bar{x}} &= -\frac{GMm}{|\bar{x}|^2}\hat{x} + m\left(\bar{\nabla}\ln\hbar\cdot\dot{\bar{x}}\right)\dot{\bar{x}} - (\bar{\nabla}\ln\hbar)\left(\frac{1}{2}m|\dot{\bar{x}}|^2 + \frac{GMm}{|\bar{x}|}\right) \\
m\ddot{r} = F &= -\frac{GMm}{r^2} + \partial_r \ln\hbar\left(\frac{1}{2}m\dot{r}^2 - \frac{GMm}{r}\right) \\
r_o &= -\frac{1}{\partial_r \ln\hbar(r_o)} > 0 \\
\frac{v_o^2}{2} &= -\frac{1}{\partial_r \ln\hbar}\frac{GMm}{r^2} - \frac{GMm}{r}
\end{aligned} \qquad \text{(112a-d)}$$

where (112b) is for the situation with no velocity other than radial, and a radially dependent $\hbar$. It is possible now for the particle to remain at rest in the gravitational field, which is normally not possible except infinitely far away from $M$. It will be so if placed at zero velocity at a radius equal to (112c). For Planck's constant profiles that decrease monotonically with radius, it can be shown that $r_o = 0$ and $\infty$ are the only values admitted for (112d). There is also a tangential velocity to the gradient at which the total radial force goes to zero, $v_o$, Equation (112d).

### 3.5 Model Parameterization at Different Scales

The NGDP model offers a new degree of freedom that may be applied to behavior at varying length scales. There may be interactions between the behaviors at varying length scales, where that of a longer length scale sets an overall trend that the smaller length scale behavior is superimposed on.

#### 3.5.1 Galactic Scale Parameters: Galaxy Rotation Curves and Dark Matter

Can a position dependent Planck's constant, in and of itself, explain away dark matter, even if restricted to the context of only galaxy rotation curves, ignoring all other phenomena explained by it (BAO, CMB, anisotropies, Inflation, etc.), while also not exceeding any astrophysical or cosmological constraints? The answer is fully negatory: For NGDP, there would *not* be a MOG-, TeVeS-, or MoND-like explanation of the rotation curves without dark matter.

It is possible to find a profile of $\hbar$ that would lead to the flattening of a galaxy rotation curve without dark matter with NGDP, although it requires a large variation in $\hbar$, and this will be shown. From (109a) for a radially dependent $\hbar$, a general radially dependent classical potential energy $V_c$ and a velocity $v$ perpendicular to the gradient, the total force is radial. $V_c$ is general, and it can contain the potential energy of multiple distributions of matter (luminous, dark, and point-like). The term $v_n$ is the net velocity from all matter, and $v$ is the measured velocity. Set equal to the centripetal force for a circular orbit, the velocity may be solved for, producing (113a-d). Setting the velocity to be a constant independent of radius per a perfectly flat rotation curve of infinite range, one may solve for the profile using (113a-d). To find values to insert, the rotation curve is computed for a dark matter containing galaxy, using the values of $M$ = $1.3\times10^{11}$ solar masses for all the visible stars in the galaxy, and a dark matter halo of $1.5\times10^{12}$ solar masses, where at a distance of 60 kpc the velocity is about 150 km/s, and flattening out. No point mass for the black hole is used, as it does not impact the rotation curve in the outer points of the galaxy. Then, using the latter values for $v$, and $M$ with $\varphi_c = -GM/r$ but without the mass of the dark matter, it is found that in order to maintain the constant velocity, $\hbar$ would have to decrease by a relative factor of unity to 0.755 over 10-60 kpc reaching a minimum mid-range, using (113d). The required mid-range minimum becomes a factor of 0.9 for $M = 9\times10^{10}$ solar masses over 10-30 kpc. This is the simplest approximation for a star near the edge of the galaxy, illustrating that a minimum can result.

While more realistic simulations with actual non-dark matter density profiles and real rotation curves may reduce the required change in $\hbar$, it is not likely to be small, and would be problematic for fusion in most of the stars in the flat velocity region, therefore, violating an astrophysical constraint.

$$
\begin{aligned}
v^2(r) &= \frac{-\partial_r V_c + V_c \partial_r \ln\hbar}{-\dfrac{m}{r} + \dfrac{m}{2}\partial_r \ln\hbar} \\
\partial_r \varphi_c &= \frac{v_n^2(r)}{r} \\
\varphi_c &= \int \frac{v_n^2(r)}{r} dr + \varphi_o \leq 0 \\
\partial_r \ln\hbar &= \frac{\dfrac{v^2(r)}{r} - \partial_r \varphi_c}{\dfrac{1}{2} v^2(r) - \varphi_c}
\end{aligned}
\qquad (113a\text{-}d)
$$

### 3.5.2 Cosmological Scale Parameters: Newtonian Derivation of Friedmann Equation for Dominant Path and the Cosmological Constant

A connection of the path integral derivation above has not been made with general relativity. In the absence of a higher theory for a total frequency-conserving version of the Einstein field equations, it is possible to derive an expression for the Friedmann equation using Newtonian gravity, following a procedure outlined in introductory texts on cosmology, adapted to include features of NGDP. Since the Lagrangian $L_c/\hbar$ is in terms of frequency and not energy, the Hamiltonian is also, is total frequency conserving as there is no explicit time dependence, and was in (108) called $W$,

$$W = \frac{H_c(r)}{\hbar(r)} = \frac{1}{\hbar(r)}\left(\frac{1}{2}m\dot{r}^2 - \frac{GMm}{r}\right) = \frac{1}{\hbar(r)}\left(\frac{1}{2}m\dot{r}^2 - \frac{4\pi G\rho r^2}{3}\right) \tag{114}$$

Equation (114) describes the frequency of a particle of mass $m$ at a radius $r$ from the origin of a homogeneous mass distribution of density $\rho$. The usual procedure is to change to the co-moving coordinates $x$ in terms of the scale factor $a$,

$$r(t) = a(t)x \tag{115}$$

Making this substitution, multiplying both sides by $2/ma^2x^2$, it is found that,

$$\left(\frac{\dot{a}}{a}\right)^2 = \frac{8\pi G}{3}\rho + \frac{2W}{ma^2x^2}\hbar(ax) \tag{116}$$

The second term of (116) must be independent of $x$ in order to maintain homogeneity. There are a number of ways this might be accomplished. One way will be examined that produces a term like the cosmological constant. In the usual derivation, the energy of a particle is constant, but changes with separations as $U \propto x^2$, to allow a connection to be made to the curvature $k$, and to arrive at the same form of expression derived from general relativity. Now, let it be assumed this persists in the conserved frequency as $W \propto x^2$, and that the variation of $\hbar$ has some unknown dependence on $ax$ shown in the middle expression in (117).

$$\hbar = \hbar_o + f_\hbar(ax) = \hbar_o + b(ax)^2 \tag{117}$$

$$\left(\frac{\dot{a}}{a}\right)^2 = \frac{8\pi G}{3}\rho + \frac{2W\hbar_o}{ma^2x^2} + \frac{2Wf_\hbar(ax)}{ma^2x^2} \tag{118}$$

Rewriting (118) using $kc^2 = -2W\hbar_o/mx^2$,

$$\left(\frac{\dot{a}}{a}\right)^2 = \frac{8\pi G}{3}\rho - \frac{kc^2}{a^2} - \frac{kc^2}{\hbar_o}\frac{f_\hbar(ax)}{a^2} \tag{119}$$

It is seen from (119) that another term enters the usual Friedmann equation. Provided that $f_\hbar(ax)/a^2$ is sufficiently constant and negative, the additional term could serve as the cosmological constant.

The last term of (119) is exactly constant if the second expression in (117) holds. From (117) the universe is still isotropic, as a quadratic change in $\hbar$ is seen in every direction. As there is no single origin of $x$, the universe is still homogeneous in the sense that at one specific position an observer concludes that $\hbar$ is multi-valued, seeing the same distribution of $\hbar$ values from every other position when treated as the origin. The issue is not homogeneity, but the multiple values. In light of Appendix 3, the problem is also mitigated here, if what one actually observes is the average of all possible values. Averaging (117) and (119) over the co-moving coordinates, only the last term is affected by the averaging, from which follows,

$$\begin{aligned}
&\langle x^2\rangle_x = \frac{\pi^2-4}{2}x_R^2 \sim \frac{\pi^2-4}{2}\frac{1}{k} \\
&\langle \hbar\rangle_x - \hbar_o = \langle f_\hbar(ax)\rangle_x = \langle ba^2x^2\rangle_x = ba^2\langle x^2\rangle_x \sim \frac{\pi^2-4}{2}\frac{ba^2}{k} \\
&\frac{\Lambda}{3} = -\frac{kc^2}{\hbar_o}b\langle x^2\rangle_x \sim -\frac{\pi^2-4}{2}\frac{bc^2}{\hbar_o} \\
&\hbar \sim \hbar_o\left(1 - \frac{2\Lambda}{3\left(\pi^2-4\right)c^2}r^2\right) \\
&\langle \hbar\rangle_x = \hbar_o\left(1 - \frac{\Lambda}{3}\frac{a^2}{kc^2}\right) \qquad \text{(120a-h)}
\end{aligned}$$

$$\begin{aligned}
&\Lambda = 9.95\times10^{-36}[1/s^2] \\
&|b| \le 1.324\times10^{-87}[Js/m^2] \\
&r_o \sim 2.822\times10^{26}[m]
\end{aligned}$$

From the second term of (120c), in order for there to be a non-zero and positive cosmological constant, $k\neq0$ so that the universe could not be *perfectly* flat, but instead open or closed. If it is closed, $k>0$, the average value of $x^2$ over $x$ would be finite and also be a constant, and then necessarily $b<0$. The theory of the inflationary universe driving it to *perfect* flatness, is in conflict with these ideas.

To go farther than this, another identification must be made. Evaluating the volume average (120a) out to a radius $x_R$, there is a resulting factor of $x_R^2$, and so the average of $x^2$ is identified as being proportional to $1/k$, from which the last terms of (120a-c) are derived. The latter is sensible only for $k \geq 0$. The geometrical factor that results will be used for the order of magnitude estimates, though (120e) is independent of it. One sees explicitly from the last term of (120b), to avoid a divergence, the factor $b$ must be zero if the universe is perfectly flat with $k$=0, causing the cosmological constant to be zero seen from (120c). Therefore, $b<0$, the Cosmological Constant is positive, and from the last term of (120e), the $\hbar$ that is measured falls with time, and apparently, can eventually become zero, or negative. The latter problem is mitigated below, as well.

The measured value $<\hbar>_x \leq \hbar_o$, so an upper limit on the absolute value of $b$ can be estimated from (120c) using the measured value of today. From (120d) the separation $r_o$ is that where a Planck's constant contribution from a distant point would fall to zero *before averaging* - remarkably on the same order as the radius of the present-day observable universe.

The latter result is interesting, even though the derivation is not rigorous, since the contribution to the average Planck's constant one then experiences would be on the order of the distances still in causal connection with the observer, and, this average value of Planck's constant would then be finite and positive. However, under these conditions, the universe would have to be ever-so-slightly closed, in conflict with most the accepted cosmological theory, Inflation, that it should approach perfect flatness, though there is still uncertainty in the measurements of the energy density of the universe, today: WMAP and Planck giving a total density parameter of 1.00±0.02; BOOMERang 1.0±0.12 [45-46]. In addition, there is no experimental constraint on the sign of the curvature at this time. An accelerated expansion is still admitted by (119). The author also points out that for the interpretation given in Section 2.7.1, a "sufficiently flat", but closed curvature was also needed.

### 3.5.3 Impact on Gravitational Radiation

An orbit for a Hulse-Talyor-like binary for a highly exaggerated effect is shown in Figure 3. Figure 3 (Left) is the normal Newtonian result without a position dependent Planck's constant, Figure 3 (Right) is NGDP using a profile per (123) of Section 4.0 with, $b_4/b_3 = 2.65\times10^8$ m. The $b_4/b_3$ value allows the Planck's constant value to vary as sensed by the other body by about 10% over the orbit. A large negative apsidal precession and orbital period increase occurs, from 8.1 to 11.78 hours.

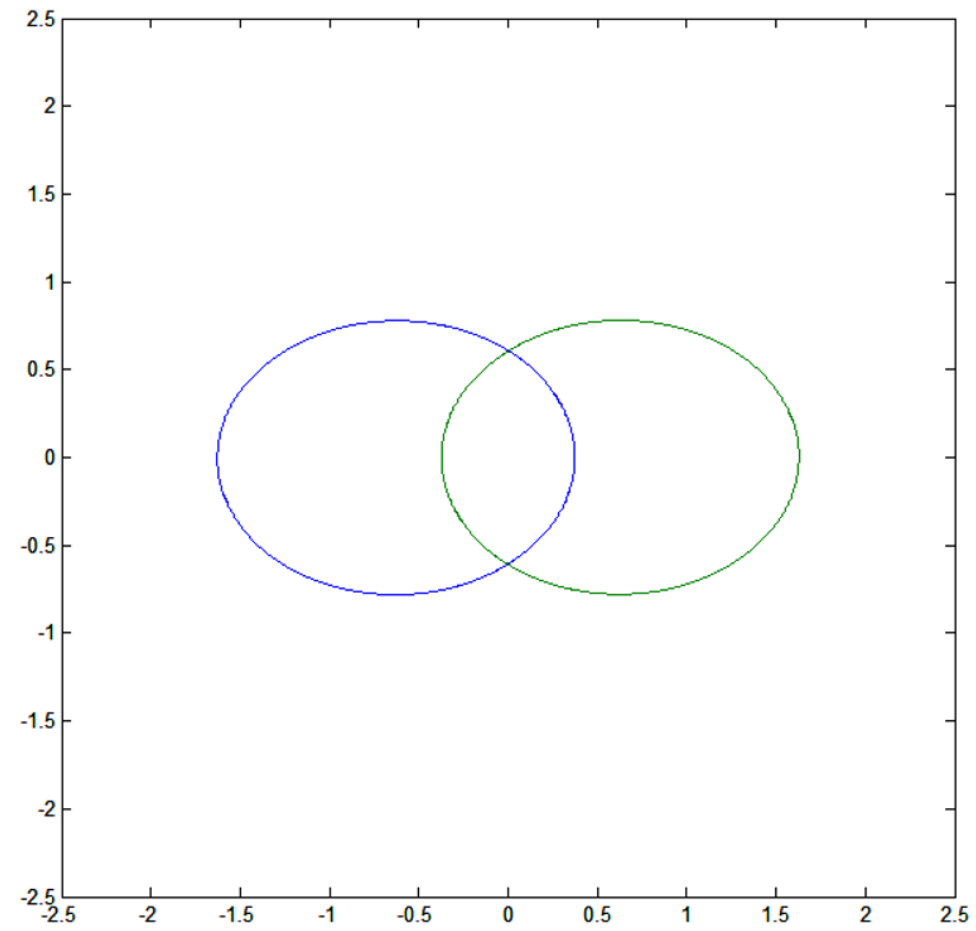

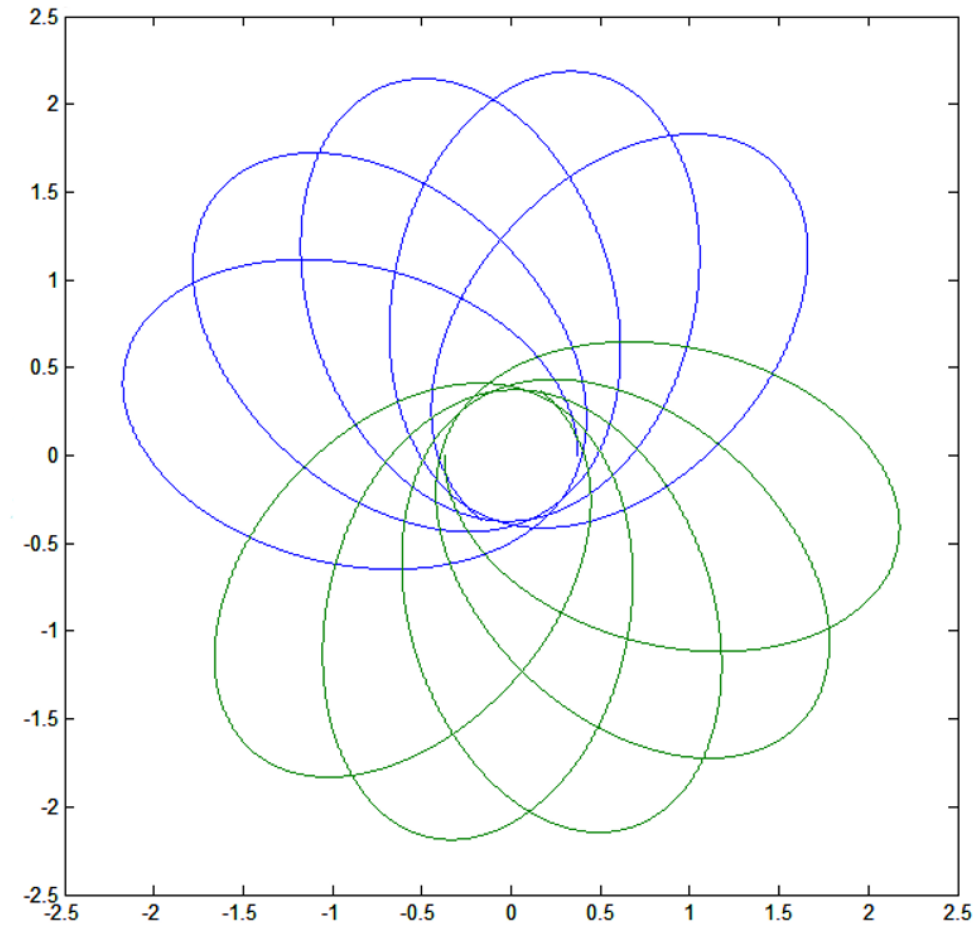

**Figure 3.** Use of Equation (109a) to numerically compute the orbit of a Hulse-Taylor-like binary. Five orbital periods are shown. Each mass is 1.4 solar masses placed at 746,000 km at periastron and launched in opposite directions vertically at 450 km/s. **(Left)** The Newtonian result is shown, with a period of 8.1 hours. There is no spatial variation in Planck's constant. **(Right)** The NGDP result, using $b_4/b_3 = 2.65\times10^8$ m, and now has a period of 11.78 hours, and a negative apsidal precession (opposite the orbital direction) of 104,000 arcseconds per period. The axis units are billions of meters.

For the example of the binary orbit being considered, its rate of period reduction by emission of gravitational waves should be per [47] with $e$=0.62 and Equation (121), derived without a variable Planck's constant, here, treated as if still valid for a sufficiently small variation,

$$\dot{P}_{GW} = -\frac{192\pi G^{5/3}}{5c^5}\left(\frac{P}{2\pi}\right)^{-5/3}(1-e^2)^{-7/2}\left(1+\frac{73}{24}e^2+\frac{37}{96}e^4\right)\frac{m_p m_c}{\left(m_p+m_c\right)^{-1/3}} \tag{121}$$

the rate of orbital period reduction is $-2.25\times10^{-12}$ s/s for the binary of normal gravity, and $-1.20\times10^{-12}$ s/s for the NGDP model with variable Planck's constant.

Therefore, for positive/negative $b_4/b_3$ the impact of such an additional force on the radiation of gravitational waves is to cause a negative/positive apsidal precession that would not normally occur, and to have larger/smaller than expected orbital periods, resulting in slower/faster than expected orbital period decrease rate, due to decreased/increased gravitational radiation emission.

### 4.0 Discussion

From [25], now reproduced *verbatim*:

"Field Theory and General Relativity are the cornerstones of modern physics. There seem to be some inherent contradictions in both theories. For example, in field theory, a static field functions much like the fields as envisioned by Faraday. Yet, a static field can be approximated with the tree-level terms of the perturbative expansion to produce an amplitude, with Feynman diagrams showing particle exchange limiting the interaction to the speed of light, equated to the Born approximation amplitude to produce a classical potential. Propagation of a field would therefore appear to be required for the static field to function. For a black hole, the mass, charge, and angular momentum are not censored: they are communicated by non-propagating modes in field theory, the accepted explanation. Changes in the static fields are propagated at the speed of light, but once reestablished, are Faraday-like static fields once again, influencing instantaneously at a distance. Physical constants would be static fields, and like any static field, described by non-propagating modes, Faraday-like, influencing instantaneously at a distance. When matter or charges gravitate into the event horizon of a black hole, the initial non-propagating modes, understood to have been set up long ago, quickly readjust, to produce the new non-propagating mode. Yet, the despite the censorship of the event horizon, somehow, in the particle picture, particles (fields) must propagate from the event horizon to reset the non-propagating modes – thus emission of particles from the horizon would seem to be needed. The descriptions conflict, despite the predictive power. As to General Relativity, it is perfectly acceptable at this time, that energy and momentum are not conserved when there are dynamical changes in spacetime, although *how* the non-conservation evolves is well understood, with the conservation possible only in a locally flat frame. With a cosmological constant, the total energy of the universe increases (explained by negative pressure). The variation of physical constants throughout the universe may also constitute other acceptable violations of conservation laws."

An object called an acton has been so named to make referring to it easier. The name encompasses:

1. the classical zero-momentum non-propagating solution
2. the classical and quantum non-propagating standing wave solution
3. the standard free-field solution

Solutions 1 and 2 are thought to be coincident with regions of high energy density. Solutions 2 and 3 may facilitate temporary variations in Solution 1 by particle propagation, resetting the non-propagating mode. An averaging procedure applied to Solution 1, in Appendix 3, leads to a cosmological temporal variation of the background value of Planck's constant. Solution 2 could be quantized with a rationale requiring a periodic boundary conditions occurring in the early universe. Solution 3 was quantized by completely standard means. All three are thought to describe local perturbations in the present epoch, about massive objects whose energy density is persistent.

Solutions for a supported scalar field were found using the zero-momentum solution of the supporting field when far from the origin of the acton. The supported field solution acquires an exponential decay beyond the free field, where the exponential decay constant is proportional to the time evolution parameter of the supporting field. If Planck's constant is increasing/decreasing slowly, then the amplitudes of supported fields decay/grow with time. Close to the origin, the behavior is more complicated, and requires solving the differential equations numerically.

The supporting field solution resembles the position dependent Planck's constant arrived at in reference [25] using a general relativistic argument, shown in (122), to which can be compared the result from field theory of this work (123),

$$\frac{\hbar(r)}{\hbar_\infty} = \left(1 - \beta_h \frac{R_S}{r}\right)^{1/2} \qquad (122)$$

$$\frac{\underline{\hbar}_{o\psi}(r,t)}{\underline{\hbar}_{o\psi}(\infty,t)} = \left(1 + \frac{b_4 / b_3}{r}\right)^{1/2} \tag{123}$$

In (122), $\beta_h$ is the Local Position Invariance (LPI) violation parameter for Planck's constant, $R_S$ is the Schwarzschild radius. The expressions come from completely different starting points. Comparing (122-123), one is prompted to conclude the origin of the acton "coincides" with the origin of a massive body. Using the mass of the sun and (122), and the value of $b_4/b_3$ from the calibration section 2.8 and died experiment reference [19] in (123), one finds $\beta_h = -5.43\times10^4$, a huge number, larger than any LPI violation ever measured, or expected. The acton may coincide with mass, but $b_4/b_3$ would be concluded not to be dependent on mass-energy, at least with the presently available information used for the calibration. If one compares $k_\alpha+0.17k_e = -(3.5\pm6)\times10^{-7}$ bounding the fine structure constant variation on the Earth about the sun from [29], the diode result [19] is enormous, suggesting a different effect, unaccounted for amplification, or systematics, where a possible reason is discussed in Appendix 1. This data is available for independent analysis from the author of [19]. Berengut's estimation from white dwarf G191 B2B was $k_\alpha = (0.7\pm0.3)\times10^{-7}$ based on Fe V [27], while from atomic clocks on Earth it is $k_\alpha = -(5.5\pm5.2)\times10^{-7}$ [48].

The standing spherical wave solution is the most easily related to early quantum cosmology. It spatially decays to zero far from its origin. It could be readily quantized, showing energy quanta similar in form to a free field, only if positive radial momentum states are allowed. To try to rationalize this requirement, the traveling waves that comprise the standing wave were interpreted to circulate continually through an infinitude of origins (to be consistent with the cosmological principle), to infinity and back around to the other side, due to imposed periodic boundary conditions from closed curvature. The field is postulated to have been set up in the $10^{-36}$ s dwell time after the Big Bang, and prior to Inflation, when and if the Universe had closed curvature, before being driven flat by Inflation. Neither expansion, or closed curvature is accounted for directly in the equations from which the result was derived, since they were set up without coupling to gravity.

The path integral result for a position dependent Planck's constant, the cause of the position dependence not being specific, was derived from a completely different starting point, from the ideas presented in [25]. The form of the path integral suggests that a variation in Planck's constant is related to the rate of time passage experienced by an object along its classical path, and this leads to unusual, additional forces. Both the cosmological interpretation of the field and the path integral hint at the need for an every-so-slight closed curvature to presently exist.

## 5.0 Conclusions

This a fertile area of research related to inflation, the inflaton, the cosmological constant, the origin of physical constants, quintessence, *k*-essense, accelerated expansion, "dark entities", CMB anisotropy, and the possible relationships between all of them.

It would be worthwhile to determine if there are any mathematical studies of fields that have *not* yet been directly correlated to variations in the physical constants most familiar us. To find these examples, a significant undertaking in itself, would result in much shorter new papers, focusing on the interpretations and implications of the existing solutions in this light. The uncorrelated work, having not been focused on the objective of understanding constants, may be especially difficult to recognize, and the setups steered towards final solution sets that are unrecastable. The author can attest to the rarity of finding literature discussing variations of Planck's constant directly. Masters of field theory are encouraged to examine whether the more exotic fields that described the early universe also evolved into the constants of today. For any fundamental constant, direct and open theories concerning their variation would lead to a larger scope of experiments aimed at validating predictions. Such experiments do not have to be astrophysical or cosmological, the present focus of most work on the subject. Existing cosmological theories, newly constant-correlated, would benefit.

As to the path integral results, methods to test the model's concrete predictions would be: 1) monitoring of gravitational wave emission of mergers that do not match expected theory; 2) monitoring the evolution of Hulse-Taylor-like binaries, showing longer/slower periods than the masses involved predict, and slower/faster orbital period reductions; 3) observation of unexpected negative/positive apsidal precession in any bound orbit.

None of these experimental verifications are necessarily easy, all may not be executable or detectable with existing technology, or performable within a single human lifetime. For example, for a Hulse-Taylor-like binary, the binaries themselves need to be found, and observed for potentially many human generations. Presently, one must infer the masses of the bodies involved from distant observations of specific orbital characteristics, and one must rely on only a self-consistency test of the model. Yet, the masses really need to be known first, as inputs, in order to test the orbital models. There must therefore be separate measurements of the mass from other kinds of observations (multi-messenger astronomy). In some cases, it may be required to at least be in closer proximity to the objects to be conclusive – the experimentalists must go there, and at this time, they cannot.

Finally, the author restates that this effort is not driven by any belief in his own theory, or even in any of the experimental data used as evidence, admittedly in scant abundance at this time, but growing, the most relevant being the spectroscopic analysis of white dwarfs looking for variations in the fine structure constant with gravitational redshift. The authors only belief is that we do not know what fundamental constants are. If they are found to behave like dynamical fields, we have a ready-made description going back to Pascual Jordan. If they do not, or are not seen to vary, then we do not have a description. The experiments will make this determination.

## 6.0 Acknowledgements

A version of this manuscript has been released as a pre-print at arXiv, ([Dannenberg]), https://arxiv.org/abs/1812.02325 , and is reference [53].

## 7.0 References

[1] P. A. M. Dirac, A New Basis for Cosmology. Proc. Royal Soc. London A 165 (921) 199–208 (1938)

[2] A. P. Meshik, The Workings of an Ancient Nuclear Reactor, Scientific American, January 26, (2009)

[3] J.-P. Uzan and B. Leclercq, The Natural Laws of the Universe: Understanding Fundamental Constants. Springer Science & Business Media, (2010)

[4] J. K. Webb, M. T. Murphy, V. V. Flambaum, V. A. Dzuba, J. D. Barrow, C. W. Churchill, and A. M. Wolfe, Further evidence for cosmological evolution of the fine structure constant. Phys. Rev. Lett. 87(9), 091301 (2001)

[5] Sze-Shiang Feng, Mu-Lin Yan, Implication of Spatial and Temporal Variations of the Fine-Structure Constant, Int. J. Theor. Phys. 55, 1049–1083 (2016)

[6] L. Kraiselburd, S. J. Landau, and C. Simeone, Variation of the fine-structure constant: an update of statistical analyses with recent data. Astronomy &Astrophysics 557, (2013)

[7] G. Mangano, F. Lizzi, and A. Porzio, Inconstant Planck's constant, Int. J. of Mod. Phys. A 30(34) (2015) 1550209

[8] Maurice A. de Gosson, Mixed Quantum States with Variable Planck's Constant, Physics Letters AVolume381, Issue 36, Pages 3033-303 (2017)

[9] Jean-Philippe Uzan, The fundamental constants and their variation: observational and theoretical status, Reviews of Modern Physics, 75, 403-455 (2003)

[10] J. Kentosh and M. Mohageg, Global positioning system test of the local position invariance of Planck's constant, Phys. Rev. Lett. 108(11) (2012) 110801

[11] J. Kentosh and M. Mohageg, Testing the local position invariance of Planck's constant in general relativity. Physics Essays 28(2), 286–289 (2015)

[12] Sven Herrmann, Felix Finke, Martin Lülf, Olga Kichakova, Dirk Puetzfeld, Daniela Knickmann, Meike List, Benny Rievers, Gabriele Giorgi, Christoph Günther, HansjörgDittus, Roberto Prieto-Cerdeira, Florian Dilssner, Francisco Gonzalez, Erik Schönemann, Javier Ventura-Traveset, and Claus Lämmerzahl, Test of the Gravitational Redshift with Galileo Satellites in an Eccentric Orbit Phys. Rev. Lett. 121, 231102 (2018)

[13] Ellis, K.J., The Effective Half-Life of a Broad Beam 238 Pu/Be Total Body Neutron Radiator. Physics in Medicine and Biology, 35, 1079-1088(1990)http://dx.doi.org/10.1088/0031-9155/35/8/004

[14] Falkenberg, E.D., Radioactive Decay Caused by Neutrinos? Apeiron, 8, 32-45(2001)

[15] Alburger, D.E., Harbottle, G. and Norton, E.F., Half-Life of 32Si. Earth and Planetary Science Letters, 78, 168-176(1986). http://dx.doi.org/10.1016/0012-821X(86)90058-0

[16] Jenkins, J.H., et al., Analysis of Experiments Exhibiting Time Varying Nuclear Decay Rates: Systematic Effectsor New Physics?(2011) http://arxiv.org/abs/1106.1678

[17] Parkhomov, A.G., Researches of Alpha and Beta Radioactivity at Long-Term Observations.(2010) http://arxiv.org/abs/1004.1761

[18] Siegert, H., Schrader, H. and Schötzig, U., Half-Life Measurements of Europium Radionuclides and theLong-Term Stability of Detectors. Applied Radiation and Isotopes, 49, 1397(1998).http://dx.doi.org/10.1016/S0969-8043(97)10082-3

[19] Richard A. Hutchin, Experimental Evidence for Variability in Planck's Constant, Optics and Photonics Journal, 6, 124-137 (2016)

[20] Cooper, P.S. (2008) Searching for Modifications to the Exponential Radioactive Decay Law with the Cassini Spacecraft.http://arxiv.org/abs/0809.4248

[21] Norman, E.B., Browne, E., Chan, Y.D., Goldman, I.D., Larimer, R.-M., Lesko, K.T., Nelson, M., Wietfeldt, F.E. andZlimen, I., Half-Life of 44Ti. Physical Review C, 57 (1998)

[22] E. N. Alexeyev, Ju. M. Gavriljuk, A. M. Gangapshev, A. M. Gezhaev, V. V. Kazalov, V. V. Kuzminov, S. I. Panasenko, S. S. Ratkevich, S. P. Yakimenko, Experimental test of the time stability of the half-life of alpha-decay Po-214 nuclei, (2011) https://arxiv.org/abs/1112.4362

[23] Eric B. Norman, Additional experimental evidence against a solar influence on nuclear decay rates (2012) https://arxiv.org/abs/1208.4357

[24] Karsten Kossert, Ole Nähle, Disproof of solar influence on the decay rates of 90Sr/90Y (2014) https://arxiv.org/abs/1407.2493

[25] Rand Dannenberg, Position Dependent Planck's Constant in a Frequency-Conserving Schrödinger Equation, *Symmetry* (2020), 12, 490; doi:10.3390/sym12040490 https://www.mdpi.com/2073-8994/12/4/490

[26] J. Hu, J. K. Webb, T. R. Ayres, M. B. Bainbridge, J. D. Barrow, M. A. Barstow, J. C. Berengut, R. F. Carswell, V. Dumont, V. Dzuba, V. V. Flambaum, C. C. Lee, N. Reindl, S. P. Preval, W.-Ü. L. Tchang-Brillet, Measuring the fine structure constant on a white dwarf surface; a detailed analysis of Fe V absorption in G191-B2B, (2020) https://arxiv.org/abs/2007.10905

[27] Berengut, J.C.; Flambaum, V.V.; Ong, A.; Webb, J.K.; Barrow, J.D.; Barstow, M.A.; Holberg, J.B. Limits on the dependence of the fine-structure constant on gravitational, potential from white-dwarf spectra. Phys.Rev. Lett. (2013), 111, 010801.

[28] J. Bagdonaite, E. J. Salumbides, S. P. Preval,M. A. Barstow, J. D. Barrow,M. T. Murphy,and W. Ubachs› Limits on a Gravitational Field Dependence of the Proton–Electron Mass Ratio from H2 in White Dwarf Stars, 10.1103/PhysRevLett.113.123002 (2014) , https://arxiv.org/pdf/1409.1000.pdf

[29] V. V. Flambaum and E. V. Shuryak, How changing physical constants and violation of local position invariance may occur?, AIP Conf. Proc. 995, 1 (2008), arXiv:physics/0701220.

[30] C. Wetterich, Naturalness of exponential cosmon potentials and the cosmological constant problem (2018), preprint https://arxiv.org/abs/0801.3208

[31] C. Wetterich, Cosmon inflation. (2013) preprint https://arxiv.org/abs/1303.4700

[32] J.D.Bekenstein, Fine-structure constant: Is it really a constant, Phys. Rev. D 25,1527 (1982).

[33] J.D. Bekenstein, Fine-structure constant variability, equivalence principle and cosmology, Phys. Rev. D66, 123514 (2002).

[34] A. Albrecht; J. Magueijo, A time varying speed of light as a solution to cosmological puzzles. Phys. Rev. D59 (4): 043516 (1999).

[35] J.D. Barrow, Cosmologies with varying light-speed. Physical Review D. 59 (4): 043515 (1998)

[36] J. W. Moffat, Superluminary universe: A Possible solution to the initial value problem in cosmology Int.J.Mod.Phys. D2 351-366 (1993)

[37] J. W. Moffat, Variable Speed of Light Cosmology, Primordial Fluctuations and Gravitational Waves, Eur. Phys. J. C 76:13 (2016)

[38] C.R. Almeida, J.C. Fabris, F. Sbisá, and Y. Tavakoli, Quantum cosmology with k-Essence theory, proceedings of the 31st International Colloquium on Group Theoretical Methods in Physics. Rio de Janeiro, Brazil, 19-25 June 2016 https://arxiv.org/abs/1604.00624v2

[39] Alexander Vikman, K-essence: cosmology, causality and emergent geometry, Doctoral Dissertation Ludwig–Maximilians–Universität München (2007)

[40] Rub´en Cordero, Eduardo L. Gonz´alez, Alfonso Queijeiro, An equation of state for purely kinetic k-essence inspired by cosmic topological defects, (2016) https://arxiv.org/abs/1608.06540v2

[41] C. Armend´ariz-Pic´on, T. Damour, V. Mukhanov, k -Inflation, Phys.Lett.B458:209-218, (1999), https://arxiv.org/abs/hep-th/9904075v1

[42] Rong-Jia Yang, Xiang-Ting Gao, Phase-space analysis of a class of k-essence cosmology, Class.Quant.Grav.28:065012,(2011) https://arxiv.org/abs/1006.4986v3

[43] Rand Dannenberg, Excluded Volume for Flat Galaxy Rotation Curves in Newtonian Gravity and General Relativity, Symmetry 2020, 12, 398; doi:10.3390/sym12030398, https://www.mdpi.com/2073-8994/12/3/398

[44] Rand Dannenberg, Einstein-Hilbert Action And Matter Action With A Particular Scalar Field Dependence (2019), DOI: 10.13140/RG.2.2.27863.32168/20, https://www.researchgate.net/publication/333704308_Einstein-Hilbert_Action_and_a_Matter_Action_with_a_Particular_Scalar_Field_Dependence

[45] Ade, P. A. R.; Aghanim, N.; Armitage-Caplan, C.; Arnaud, M.; Ashdown, M.; Atrio-Barandela, F.; Aumont, J.; Baccigalupi, C.; Banday, A. J.; Barreiro, R. B.; Bartlett, J. G.; Battaner, E.; Benabed, K.; Benoît, A.; Benoit-Lévy, A.; Bernard, J.-P.; Bersanelli, M.; Bielewicz, P.; Bobin, J.; Bock, J. J.; Bonaldi, A.; Bond, J. R.; Borrill, J.; Bouchet, F. R.; Bridges, M.; Bucher, M.; Burigana, C.; Butler, R. C.; Calabrese, E.; et al. (2014). "Planck2013 results. XVI. Cosmological parameters". Astronomy & Astrophysics. 571: A16. arXiv:1303.5076. Bibcode:2014A&A...571A..16P. doi:10.1051/0004-6361/201321591.

[46] De Bernardis, P.; Ade, P. A. R.; Bock, J. J.; Bond, J. R.; Borrill, J.; Boscaleri, A.; Coble, K.; Crill, B. P.; De Gasperis, G.; Farese, P. C.; Ferreira, P. G.; Ganga, K.; Giacometti, M.; Hivon, E.; Hristov, V. V.; Iacoangeli, A.; Jaffe, A. H.; Lange, A. E.; Martinis, L.; Masi, S.; Mason, P. V.; Mauskopf, P. D.; Melchiorri, A.; Miglio, L.; Montroy, T.; Netterfield, C. B.; Pascale, E.; Piacentini, F.; Pogosyan, D.; et al. (2000). "A flat Universe from high-resolution maps of the cosmic microwave background radiation". Nature. 404 (6781): 955–9. arXiv:astro-ph/0004404. Bibcode:2000Natur.404..955D. doi:10.1038/35010035. PMID 10801117.

[47] Joel M. Weisberg, Joseph H. Taylor, The Relativistic Binary Pulsar B1913+16: Thirty Years of Observations and Analysis, Binary Radio Pulsars ASP Conference Series, Vol. 328, (2005)

[48] N. Leefer, C. T. M. Weber, A. Cingo ̈z, J. R. Torg- erson, and D. Budker, “New limits on variation of the fine-structure constant using atomic dysprosium”, arXiv:1304.6940 (2013).

[49] M.J. Duff, Comment on time-variation of fundamental constants, (2002), https://arxiv.org/abs/hep-th/0208093

[50] M.J. Duff, How fundamental are fundamental constants? (2014) https://arxiv.org/abs/1412.2040, DOI 10.1080/00107514.2014.980093

[51] M. J. Duff, L. B. Okun, G. Veneziano, Trialogue on the number of fundamental constants, (2002), https://arxiv.org/abs/physics/0110060, DOI 10.1088/1126-6708/2002/03/023

[52] Zhe Chang, Qing-Hua Zhu, Spatial variation of the fine-structure constant and Hubble's law in anisotropic coordinate of Friedmann-Lemaitre-Robertson-Walker space-time, https://arxiv.org/abs/2011.07773 (2020)

[53] Rand Dannenberg, Planck’s Constant As A Dynamical Field & Path Integral (2020), arXiv:1812.02325 [quant-ph], https://arxiv.org/abs/1812.02325

**Appendix 1**

The utility/futility of measuring dimensionful versus dimensionless fundamental constant variations is still a controversial subject, undecided in the literature [49-51], with published differences in opinion. The fine structure constant is an example of the former type, as are ratios of particle masses, coupling constants of fundamental forces, while *c*, *e*, *h* are examples of the latter. The references given contain dispute even over the number of fundamental constants that there are, and whether there are different physical consequences for variations of the dimensioned ones.

One may write a theory for a dimensioned constant, and another for a dimensionless one, and since there is no undisputed data in the literature concerning variations of either type, any reasoning here is equivalent to opinion, a recent one may be found here [52]. This author will offer his own.

There are two types of systematics influencing the instrumentation being used to make a measurement of the varying constant:

1. The *usual* sort, in which a progressively lower or higher measurement of a length, mass, or time is progressively higher or lower from the truth, respectively, due to undiscovered non-linearity or a calibration problem with the instrument(s).
2. The *unusual* sort in which the field of a fundamental "constant" is changing, causing alterations of the instrument(s) response in some manner.

As an example, consider the following test concerning the speed of light. An experimentalist, *A*, is making measurements of the speed of light with a light-source, detector, and light-clock, several meter sticks, all made of different materials, situated in a laboratory that is in freefall with him, towards the event horizon of a black hole. Experimentalist *B* remains very far away from the horizon, with an identical laboratory.

*A* is subject to both forms of systematics, while *B* must contend with only the usual sort. However, the *usual* systematics can in principle be eliminated by careful calibration infinitely far away from the horizon, and this is assumed now to be so.

The *unusual* systematics encountered in *A*'s frame could be eliminated with a model for how the instrument(s) response should change given a variation of the constant under study, but it will be assumed for now this model does not exist.

When *A* was next to *B*, far away from the event horizon, they together measured the speed of light to have a value $c_o$.

Neither *A* nor *B* know for certain whether the speed of light is a dynamical field, but they have seen the equations for when any constant is elevated to a field that is coupled to gravity, the field is $c=c_o\eta$, and when gravity is weak or the coupling is weak, will go as $(c(r) - c_o)/c_o = kR_s/r$. The field $\eta$ has no dimensions, and if the speed of light physically changes in *A*'s frame, it has nothing to do with the change of time, length, or mass (in the context of GR) between frame *A* or *B*, as $\eta$ is the field changing. The physical change in the speed of light may differ from the results of *A*'s choice of measurement technique.

If $c$ is truly constant per GR as it is presently understood, $A$ observes his light-clock ticking as per the usual, and he himself experiences no sense of change of the progression of his life. His different-material meter sticks look the same length, fitting perfectly at all times between the source-detector setup, as they fall towards the event horizon.

Suppose that $c$ really is coupled to gravity per the former paragraph. When $A$, closer to the horizon, makes his measurement, he finds a value different than $c_o$ depending both on $k$, and the *unusual* systematics of his instrument(s).

One then asks whether this measurement is "useful", or not. One will note in the following a distinction between "useful" and "truth". It is certainly "useful" for determining whether there has been a violation of local position invariance, a significant finding. Without a model for the influence of the *unusual* systematics on the instrument(s), one can also compare the measured value found by $A$ at coordinate $r$ to what it should have been from the theory independent of the *unusual* systematics, to see if they differ significantly, or not. This is "useful", because it bounds the *unusual* systematics.

The field of $c$ is "infused" in the atomistic and electronic structure of $A$'s meter sticks, himself, his light-clock, detector apparatus, all of his instrumentation, supporting electronic devices, computers, all of the different materials involved, etc.

Consider only $A$'s meter sticks, as an example. He has one made of a light elements (H,C,N,O) whose electronic structure are very well described in the solid state by Density Functional Theory (DFT) without relativistic effects (atomic number to speed of light ratio $Z/c$ is small). He has a second made of heavy elements (Cs, Pt, Au, Pb, U) where the relativistic effects become important ($Z/c$ is large). They were cut to exactly the same length infinitely far from the horizon, and ruled. A changing value of $c$ as an input would result in no change in length of the former, but some change in length of the latter, whose ratio may be read (in principle) off the rule and compared by $A$ to extract the change in $c$, based on pre-computing the ratios as a function of $c$. This is not exactly equivalent to the measurement of a dimensionless constant, because the length ratio comparison results in a measurement of a dimensioned one. Further, the length ratio magnitude as a function of $r$ does not equal the gravity-to-$\eta$ coupling magnitude, they are virtually unrelated, though may trend together.

One may then compare the value of $c$ determined by the meter stick comparisons of different materials to the value determined by the light propagation, to find whether the *unusual* systematics are of any consequence. In the light propagation test, there is of course a dilemma in measuring the propagation distance, owing to the non-agreement between the metersticks. Did the different meterstick lengths and different results using them for $c$ lead to a significant amplification or diminishment from "truth". This quantifies the *unusual* systematics - very "useful". Note that "truth" is derived from a dimensionless length ratio.

This author notes again the large value of the local position invariance violation parameter deduced from the diode experiment for Planck's constant [19] in Section 4.0 and Equations (122) and (123).

These above arguments apply even to constants that are not dimensioned, measured in the local free-falling laboratory. One cannot guarantee that the *unusual* systematic influence of the "constant" field on the multiple, differing, often extremely complex instruments measuring it will cancel out, only because the constant measured was dimensionless. The cancellation effect occurs only in principle, with idealized equipment, in which all measurements of time, length, mass, with said multiple instruments, are affected by the same three factors among the instruments, respectively. The latter is a lot to hope for, given that all the different materials comprising the instruments will have different responses to the changing $c$.

Whether the constant is dimensioned or dimensionless, a deduced variation from $c_o$ in a local experiment by $A$ is an "useful" result, even if it is quantitatively dominated by the *unusual* systematics of the instrumentation response to the field of the "constant". What one needs for good quantitative results in the local frame is a model of the *unusual* systematic instrument responses (and materials) to the changing field of the "constant", that is, "truth".

Now consider $B$, observing $A$'s lab remotely. His instruments are free of the *unusual* systematic effects caused by field he is attempting to measure, so the hoped-for, perfect cancellation effect of measuring dimensionless constants is not something he need be concerned about. So, remotely measured changes in dimensioned or dimensionless constants by $B$ are "useful", as well.

Anything that is "useful" is also "meaningful".

As an aside, assuming the *unusual* systematics are not of significant consequence, as the speed of light changes in $A$'s frame, in order for $A$ to have no perception of a change in the ticking of his light-clock, relative to his heart rate, breathing, or eye-blinking, the frequency of light must be conserved, while the wavelength varies, looking bluer or redder to $A$. Frequency conservation is treated in [25], and Sections 2.3 and all of 3.0 in this paper.

**Appendix 2**

From (26) and (27) for $p=p_r\neq 0$, a more general alternative solution is,

$$S_{p\chi}(r) = u_1 e^{-ipr} - \frac{iu_2 e^{ipr}}{2pr} \tag{A1}$$

$$\varphi_{p\chi}(x_o) = u_3 \sin\left(px_o\right) + u_4 \cos\left(px_o\right) \tag{A2}$$

However, in this solution $\chi_p \neq \psi_p^2$ and there would be additional cross terms in the construction of the full $\chi$ when squaring the sum of the $\psi_p$.

Setting $u_1 = u_3 = 0$, $u_2 = 2i(d_2)^2$, $u_4 = (d_4)^2$, and $p \rightarrow (2)^{1/2}p$, one finds,

$$\chi_p = \left(d_2 d_4\right)^2 \frac{\exp\left(i\sqrt{2}pr\right)}{\sqrt{2}pr} \cos\left(\sqrt{2}px_o\right) \tag{A3}$$

Taking the real part of (A3) results in (47) for $d_1$=$d_3$=0, reproducing the special case when $\chi_p = \psi_p^2$, and the solutions can be summed (integrated).

An additional unstudied solution is therefore also constituted by (A1) and (A2), for which $\chi_p \neq \psi_p^2$.

**Appendix 3**

Consider solutions (35) and (36) for $\chi$. To be consistent with the Cosmological Principle, the solution must be sourced at every position in space. Averaging (36) over a spherical volume $V$ of radius $R$, and then taking the limit $R$=∞, one finds,

$$\left\langle \hbar^2 \right\rangle_V = \beta^2 \left\langle \chi \right\rangle_V = \beta^2 b_1 b_3 \left( 1 + \frac{b_2}{b_1} ct \right) = \underline{\hbar}^2_{o\psi}(\infty, t) \tag{A4}$$

The last term of (A4) was defined in (70), has only a temporal dependence, and is finite, despite the divergence of (36) prior to averaging.